\documentclass[symmetry,article,accept,moreauthors,pdftex]{Definitions/mdpi} 

\firstpage{1} 
\pubvolume{13}
\issuenum{3}
\articlenumber{379}
\pubyear{2021}
\copyrightyear{2021}
\externaleditor{Academic Editor: Abraham A. Ungar}
\datereceived{5 February 2021} 
\dateaccepted{22 February 2021} 
\datepublished{26 February 2021} 
\hreflink{https://doi.org/10.3390/sym13030379} 
\pdfoutput=2

\usepackage{hyperref} 

\continuouspages{yes}
\usepackage{booktabs} 
\usepackage{multirow}
\usepackage{soul} 
\usepackage{microtype}

\setitemize{parsep=6pt,itemsep=0pt,leftmargin=*,labelsep=5.5mm,align=parleft}  
\setenumerate{parsep=6pt,itemsep=0pt,leftmargin=*,labelsep=5.5mm,align=parleft}
\setlist[description]{itemsep=0mm}   

\Title{Geometric Justification of the Fundamental Interaction Fields for the Classical Long-Range Forces}

\TitleCitation{Geometric Justification of the Fundamental Interaction Fields for the Classical Long-Range Forces}

\Author{{Vesselin G. Gueorguiev} $^{1,2,}$*\orcidB{} and Andre Maeder $^{3}$\orcidA{} }

\AuthorNames{Vesselin G. Gueorguiev and Andre Maeder}

\AuthorCitation{{Gueorguiev, V.G.; Maeder, A.}}

\address{%
$^{1}$ \quad Institute for Advanced Physical Studies, {Sofia 1784, Bulgaria}\\
$^{2}$ \quad Ronin Institute for Independent Scholarship, 127 Haddon Pl., Montclair, NJ 07043, USA\\
$^{3}$ \quad Geneva Observatory, 
University of Geneva, chemin des Maillettes 51, CH-1290 Sauverny, Switzerland;  andre.maeder at unige.ch\\
}

\corres{{Correspondence}:Vesselin at MailAPS dot org }

\date{\today}

\abstract{Based on the principle of reparametrization invariance, the general
structure of physically relevant classical matter systems is illuminated 
within the Lagrangian framework. 
In a straightforward way, the matter Lagrangian contains
background interaction fields, such as a 1-form field analogous to
the electromagnetic vector potential and symmetric tensor for gravity.
The geometric justification of the interaction field Lagrangians for
the electromagnetic and gravitational interactions are emphasized.
The generalization to $E$-dimensional extended objects ($p$-branes)
embedded in a bulk space M is also discussed within the light of some familiar examples.
The concept of fictitious accelerations due to un-proper time parametrization is introduced, 
and its implications are discussed.
The framework naturally suggests new classical interaction fields
beyond electromagnetism and gravity. The simplest model with such
fields is analyzed and its relevance to dark matter and dark energy
phenomena on large/cosmological scales is inferred. Unusual pathological
behavior in the Newtonian limit is suggested to be a precursor of
quantum effects and of inflation-like processes at microscopic scales.
}


\begin{document}

\section{Introduction}

Probing and understanding physical reality goes through a classical
interface that shapes our thoughts as classical causality chains.
Therefore, understanding the essential mathematical constructions
in classical mechanics and classical field theory is important, even
though quantum mechanics and quantum field theory are regarded as
more fundamental than their classical counterparts. Two approaches,
the Hamiltonian and the Lagrangian, are very useful in theoretical physics 
\cite{Kilmister_1967,Goldstain_1980,Deriglazov2016,Gracia_and_Josep,Nikitin-StringTheory,Carinena_1995}.
In general, there is a transformation that relates these two approaches
-- the Legendre transform \cite{Kilmister_1967,Goldstain_1980,Deriglazov2016,Gracia_and_Josep}. 
For reparametrization-invariant models, however, there are problems in changing from Lagrangian to the Hamiltonian
approach \cite{Goldstain_1980,Deriglazov2016,Gracia_and_Josep,Nikitin-StringTheory,Rund_1966,Lanczos_1970}.

\textls[-15]{Fiber bundles provide the mathematical framework for classical mechanics,
field theory, and even quantum mechanics when viewed as a classical
field theory. Parallel transport, covariant differentiation, and gauge
symmetry are very {important structures}
 \cite{Pauli_1958} associated
with fiber bundles. When asking: ``What structures are important
to physics?'', one should also ask: ``Why one fiber bundle should
be more `physical' than another?'', ``Why does the `physical' base
manifold seem to be a four-dimensional Lorentzian manifold?''
\cite{Borstnik_and_Nielsen,vanDam_and_Ng,Sachoglu_2001},
and ``How should one construct an action integral for a given fiber bundle?'' 
\cite{Kilmister_1967,Carinena_1995,Feynman_1965,Gerjuoy_and_Rau,Rivas_2001}.
Starting with the tangent or cotangent bundle seems natural because
these bundles are related to the notion of a classical point-like
matter. Since knowledge is accrued and tested via experiments that
involve classical apparatus, the physically accessible fields should
be generated by matter and should couple to matter as well. Therefore,
the matter Lagrangian should contain the interaction fields, not their
derivatives, with which classical matter interacts \cite{Dirac_1958}.}

\textls[-15]{In what follows, {the principle of reparametrization invariance
is illustrated as a guiding principle in formulating physically relevant
models}. The symmetry of reparametrization invariance is a common
feature of many important physics models but 
{it has been often treated as an issue that needs to be resolved} 
to make reasonable predictions within each specific model. Since any model
based on a Lagrangian  can be reformulated into an equivalent 
reparametrization invariant model \cite{Goldstain_1980, 2011AmJPh..79..882D}, 
 this symmetry may be signaling an important fundamental principle. 
Thus, focusing the discussion on models with Lagrangians that possess such
reparametrization invariance does not restrict the generality of the
models considered but instead provides an important classification
of the possible physical systems and their interactions. 
In a nutshell, the {principle of reparametrization invariance} 
is like the covariance principle but 
{about the internal coordinates of the physical process under study}. }

\textls[-15]{The covariance principle is effectively related to the diffeomorphism symmetry of a manifold $M$. In the theory of manifolds, switching from one chart on $M$ to another is physically equivalent to switching from one observer to another in the spacetime of the observer $M$. Thus, coordinate independence of the physical laws and their mathematical forms when formulated in the manifold framework of $M$. However, the framework does not say anything about a specific physical process $E$ until one makes the relevant manifold model.}

A physical process $E$ can be viewed as a manifold that consists of the points involved in the process 
and their relationships. Thus, processes and their studies can be viewed as the embedding of manifolds.
{That is, how should $E$  be embedded in $M$?}
For example, the motion of a point particle is just about the trajectory of a particle
as viewed as the 1-dimensional curve in the 4D space-time of the physical observers. 
Since a process $E$ is also viewed as a manifold then there are mathematical charts that describe $E$ locally.
The description of the process (embedding $E \hookrightarrow M$) should not depend on the choice made for the charts on $E$.
{This is an additional symmetry to the covariance principle that things should not depend on the choice of the observer's coordinates for $M$.}

Thus, the mathematical framework should also possess diffeomorphisms symmetry of the manifold $E$.
That is, in the case of point particle, a 1D curve (the trajectory) mapped into another topologically equivalent 1D curve 
(but same trajectory) should not change our understanding and the description of the motion of a point particle.
Physically, one talks about the covariance principle when considering the diffeomorphisms symmetry of a manifold $M$ and
about reparametrization invariance when considering the diffeomorphisms symmetry of the manifold $E$ that is embedded in $M$.

The current research suggests that reparametrization invariance can be achieved by using 
the Lagrangian formulation with Lagrangians that are homogeneous functions of order one with respect to the velocity.
This leads to all the subsequent results that justify only electromagnetic and gravitational classical
forces at  macroscopic scales which is consistent with experimental observations.

This paper aims to illustrate the possibility that physical
reality and observed physical laws are related to a mathematical construction
guided by {{the principle of reparametrization
invariance for the embedding of manifolds}}. 
This principle suggests geometric justification of the fundamental interaction fields for
the classical long-range forces -- electromagnetism and gravity, as
well as {possible new classical fields}. Are there observable
consequences of such fields on microscopic and/or cosmological scales?
Are such fields present in nature? Under what conditions  could the relevant
reparametrization invariant Lagrangians be reduced or not  with such fields 
to the known Lagrangians that contain 
only gravitational and electromagnetic fields? 
These are only a few of the far-reaching questions related to the idea of 
reparametrization invariance and its correspondence to the observed physical laws. 
One day, hopefully, some of the readers of this paper will be able to 
address these questions fully and answer them completely. 

In brief, the paper starts with the relativistic particle 
\cite{Rund_1966,Pauli_1958,Feynman_1965,Landau_and_Lifshitz}
aiming to illustrate the main ideas and their generalization to extended
objects ($p$-branes). In answering the question: 
``What is the Lagrangian for the classical matter?'' the proposed 
{canonical matter Lagrangian} naturally contains background interaction fields,
such as a 1-form field analogous to the electromagnetic vector potential
and symmetric tensor that is usually associated with gravity. 
The guiding principles needed for the construction of the Lagrangians for the 
interaction fields are also discussed as an illustration of the uniqueness of the 
Lagrangians for the electromagnetic and gravitational fields.  
The authors consider this mathematical framework to be a geometric
justification of electromagnetism and gravity. The framework presented here 
seems to be able to go beyond the Feynman\textquoteright s proof
of the Maxwell and Lorentz equations that justify electromagnetism
from a few simple fundamental principles \cite{Dyson1990AmJPh..58..209D}. 

In Section \ref{Relativistic Particle}, the Lagrangian for a relativistic
particle is given as an example of a reparametrization-invariant action.
Section \ref{Homogeneous Lagrangians} contains arguments in favor 
of first-order homogeneous Lagrangians; in Section \ref{Pros and Cons} 
are listed some of the good and lesser
properties of such models;  in Section \ref{Canonical Form} 
the canonical form of the first-order homogeneous Lagrangians is justified; and
in Section \ref{Extended Objects} the canonical structure of 
a reparametrization invariant Lagrangian for an extended object (p-brane) 
embedded in a bulk space M is shown to lead to some familiar Lagrangians, 
such as the relativistic point particle in an electromagnetic field,
the string theory Lagrangian, and the Dirac--Nambu--Goto Lagrangian.
The outlined systems are based on first-order homogeneous Lagrangians in the 
velocity/generalized velocity to achieve reparametrization invariance along 
with the usual general covariance.
Section \ref{New Forces} discusses the physical implication of such Lagrangians, 
in particular, in Section \ref{Simplest  Sn Lagrangian Systems} 
the possibility of classical forces beyond electromagnetism and gravity is studied for 
the simplest possible Lagrangian system with symmetric fields $S_n$ with $n>2$, 
while in Section \ref{Proper Time} the notion of proper time is shown to be 
mostly related to the gravitational term ($n=2$) of the matter Lagrangian,
while in Section \ref{Un-Proper Time} the consequence of utilizing 
un-proper time parametrization of a non-reparametrization-invariant action is illustrated.
Section \ref{Field Lagrangians} justifies the field Lagrangians relevant 
for the interaction fields, in particular, the uniqueness of the Lagrangian for electromagnetism, 
as well as the uniqueness of the Hilbert--Einstein action integral for gravity. 
The conclusions and discussions are given in Section \ref{Conclusions} 
followed by Section \ref{Exercises}, which contains
relevant theorems framed as problems and exercises.

\section{The Relativistic Particle Lagrangian \label{Relativistic Particle}}

It is well known that localized particles move with a finite 3D speed. 
In an extended configuration space (4D space-time), 
when the time is added as a coordinate ($x^{0}=ct$),
particles move with a constant 4-velocity ($v\cdot v=constant$).
The 4-velocity is constant due to the definition $v^{\mu}=dx^{\mu}/d\tau$
that uses the invariance of the {proper-time} ($\tau$) mathematically
defined via the symmetric tensor $g_{\mu\nu}$ ($d\tau^{2}=g_{\mu\nu}dx^{\mu}dx^{v}$).
Physically the proper-time ($\tau$) is associated with the passing
of time measured by a co-moving clock that is at rest with respect
to the particle during its motion. In this case, the action integral
for a massive relativistic particle has a nice geometrical meaning:
{It is the time-elapsed along the particle trajectory} \cite{Pauli_1958}:
\begin{eqnarray}
S_{1}=\int d\tau L_{1}(x,v) & = & \int d\tau\sqrt{g_{\mu\nu}v^{\mu}v^{\nu}},\label{S1}\\
\quad\sqrt{g_{\mu\nu}v^{\mu}v^{\nu}}\rightarrow1 & \Rightarrow & S_{1}=\int d\tau.\nonumber 
\end{eqnarray}
However, for massless particles, such as photons, the length of the
4-velocity is zero ($g_{\mu\nu}v^{\mu}v^{\nu}=0$). Thus, one has to
use a different Lagrangian to avoid problems due to division by zero
when evaluating the Euler--Lagrange equations. In this case, the appropriate
``good'' action is \cite{Pauli_1958}: 
\begin{equation}
S_{2}=\int L_{2}(x,v)d\tau=\int g_{\mu\nu}v^{\mu}v^{\nu}d\tau.\label{S2}
\end{equation}
Notice that the Euler--Lagrange equations obtained from $S_{1}$ and
$S_{2}$ are equivalent, and both are equivalent to the geodesic equation
as well: 
\begin{eqnarray}
\frac{d}{d\tau}\vec{v}=D_{\vec{v}}\vec{v}=v^{\beta}\nabla_{\beta}\vec{v} & = & 0,\label{geodesic equations}\\
\quad v^{\beta}\left(\frac{\partial v^{\alpha}}{\partial x^{\beta}}+\Gamma_{\gamma\beta}^{\alpha}v^{\gamma}\right) & = & 0.\nonumber 
\end{eqnarray}
In General Relativity (GR), the {Levi--Civita connection} $\nabla_{\beta}$,
with Christoffel symbols $\Gamma_{\beta\gamma}^{\alpha}=g^{\alpha\rho}\left(g_{\rho\beta,\gamma}+g_{\rho\gamma,\beta}-g_{\beta\gamma,\rho}\right)/2$,
preserves the length of the vectors ($\nabla g(\vec{v},\vec{v})=0$)
\cite{Pauli_1958}. Therefore, these equivalences are not surprising
because the Lagrangians in (\ref{S1}) and (\ref{S2}) are functions
of the preserved arc length $g(\vec{v},\vec{v})=\vec{v}^{2}$. In
principle, however, the parallel transport for an arbitrary connection
$\nabla_{\beta}$ does not have to preserve the length of a general
vector \cite{Pauli_1958,Weyl:1993kh}. 
This is clearly seen when metric tensor is velocity dependent and thus the usual argument will not apply \cite{Randers41}.

Remarkably, however, going beyond length preserving parallel transport may still hold such equivalence. 
For example, the Weyl's integrable geometry does have such equivalence between what one expects to be the generalized 
geodesic equation and the equation derived from an appropriate Lagrangian \cite{1978Ap&SS..54..497B}.
Weyl's integrable geometry provides a framework that is likely to be relevant to physics \cite{1979A&A....73...82M}. 
It is based on the original Weyl's gauge symmetry idea where the length of a vector may depend 
on the gauge choice as well as upon infinitesimal local displacements. 
In Weyl's integrable geometry, however, this freedom is constrained to constructions where the 
length of a vector does not change upon a transport along a closed loop.
In such geometry, one finds that only an action that is built upon a co-scalar of order ($-1$)
results in trajectory restricting equations of motion that do correspond to the generalized 
geodesic equation \cite{1978Ap&SS..54..497B} while any other choices built upon 
co-scalar length $l$ of order $n\ne-1$  results in the statement that $dl$ is a closed 
one-form, that is, a perfect deferential ($\mathrm{d}(dl)=0$). 
In Weyl's geometry terminology a scalar, vector, and a general tensor object $Y_{\mu \dots \nu}$ 
is a co-tensor of order $n$ when $\tilde{Y}_{\mu \dots \nu}=\beta^n\,Y_{\mu \dots \nu}$ 
upon the gauge change of the metric tensor $\tilde{g}_{\mu \nu}=\beta^2\,g_{\mu \nu}$.
Thus, the line element $d\tau$ defined as usual to be $d\tau^2=g_{\mu \nu}dx^\mu\,dx^\nu$ is 
a co-scalar of order $(+1)$, then the co-tangent vector with components $v^\mu=dx^\mu/d\tau$ 
is seen as a co-vector of order ($-1$). The mathematical framework developed in  \cite{1978Ap&SS..54..497B} practically  shows that the only reasonable choice of
action for a massive particle, within the Weyl's integrable geometry framework, is given by the action integral (\ref{S1}).

The equivalence between $S_{1}$ and $S_{2}$ is very robust. Since
$L_{2}$ is a homogeneous function of order $2$ with respect to $\vec{v}$,
the corresponding Hamiltonian function ($h=v\partial L/\partial v-L$)
is equal to its Lagrangian ($h(x,v)=L_2(x,v)$). As long as
there is no explicit proper-time dependence then $L_{2}$ is conserved, and
so is the length of $\vec{v}$. Any parameter independent homogeneous Lagrangian
in $\vec{v}$ ($L_n(x,\beta v)=\beta^{n}L_n(x,v)$) of order $n\neq1$ is conserved because $h=(n-1)L_n$.
When $dL/d\tau=0$, then one can show that the Euler--Lagrange equations
for $L$ and $\tilde{L}=f\left(L\right)$ are equivalent under certain
minor restrictions on $f$ (see Section \ref{Un-Proper Time} for more details). 
This is an interesting type of equivalence that applies to homogeneous Lagrangians.
It is different from the usual equivalence $L\rightarrow\tilde{L}=L+d\Lambda/d\tau$
or the more general equivalence discussed in  \cite{Hojman_and_Harleston}.
Any solution of the Euler--Lagrange equation for $\tilde{L}=L^{\alpha},\alpha\neq~1$
would conserve $L=L_{1}$ since $\tilde{h}=(\alpha-1)L^{\alpha}$
is conserved. All these solutions are solutions of the Euler--Lagrange
equation for $L$ as well; thus $L^{\alpha}\subset L$ in the sense of their set of solutions.
In general, conservation of $L_{1}$ is not guaranteed since $L_{1}\rightarrow L_{1}+d\Lambda/d\tau$ 
is also a first-order homogeneous Lagrangian in the velocities that is equivalent to $L_{1}$. 
This suggests that there could be a choice of $\Lambda$, a ``gauge fixing'', 
such that $L_{1}+d\Lambda/d\tau$ is conserved even if $L_{1}$ is not.
However, whenever $L_{1}$ is conserved, then the corresponding equations would also be
related to the geodesic Equation (\ref{geodesic equations}) as well.
Relevant examples will be discussed in the next paragraphs but before doing so,
 hands-on readers may benefit more if they do the first four problems   
of the exercises in Section \ref{Exercises}.

The simplest example is the case $L_{1}(x,v) = m \sqrt{g_{\mu\nu}v^{\mu}v^{\nu}}$ 
and $L_{2}(x,v) =\frac{m}{2} g_{\mu\nu}v^{\mu}v^{\nu}$. 
Notice that the details of the mass multiplier are actually irrelevant in this case 
since the corresponding Euler--Lagrange Equation (\ref{Euler-Lagrange equations}) is insensitive to its value.
The mass $m$, however, comes into the picture as integral of the motion as soon as 
we consider the relevant energy-momentum dispersion relation $p_{\nu}p^{\nu}=m^2$. 
For $L_{2}$ the geometric linear momentum ($p_{\nu}=mg_{\mu\nu}v^{\mu}$) and the generalized 
linear momentum ($\pi_{\mu}=\partial{L_{2}/\partial{v^{\mu}}}$) coincide. 
While in the case of $L_{1}$,  the generalized linear momentum 
$\pi'_{\mu}=\partial{L_{1}/\partial{v^{\mu}}}=\frac{mg_{\mu\nu}v^{\mu}}{\sqrt{g_{\mu\nu}v^{\mu}v^{\nu}}}$
differs from the geometric one by a factor $\frac{m}{L_{1}}=1/\sqrt{g_{\mu\nu}v^{\mu}v^{\nu}}$. 
The two linear momenta can be made the same if this factor is forced to be equal to one, that is,
to use the usual choice of proper-time parametrization that results in $g_{\mu\nu}v^{\mu}v^{\nu}=1$.
This however, is only possible for massive particle $m\ne0$ while for massless particles it is clearly 
a contradiction with their null-geodesic equation $g_{\mu\nu}v^{\mu}v^{\nu}=0$. 
Thus, for $m\ne0$ the two Lagrangians are equivalent as long as one recognizes the use of the 
proper-time parametrization. However, there is a slight nuance here, based on the reparametrization
invariance of $S_{1}$ one can see the choice of proper-time parametrization as a matter of convenience
that results in $g_{\mu\nu}v^{\mu}v^{\nu}=1$, while for $L_{2}$ this is a matter of ``physics'' content since
$S_{2}$ does not possess reparametrization invariance. Thus, a ``clever'' choice of parametrization
that reflects the physical reality has to be imposed. After all, it is still the same  proper-time parametrization 
but it is justified after looking at the Hamiltonian function for $L_{2}$ and recognizing that such choice of
parametrization would show explicitly that the Hamiltonian function corresponds to an integral of motion.
If one uses arbitrary un-proper parametrization then one is likely to come across factious acceleration as
discussed in Section \ref{Un-Proper Time}.

As stated already, the mass multiplier is irrelevant as seen from 
the corresponding Euler--Lagrange Equation (\ref{Euler-Lagrange equations}) and therefore one can
consider $S_{2}$ only with $L_{2}(x,v) =g_{\mu\nu}v^{\mu}v^{\nu}$ and use $p_{\nu}p^{\nu}=m^2$ 
as a way of assessing the mass of a particle. This is particularly useful for particles that follow the
null-geodesic equation $g_{\mu\nu}v^{\mu}v^{\nu}=0$  while the treatment of $S_{1}$ is more complicated
\cite{Pauli_1958}. The problem with $S_{1}$  steams from the fact that 
now the generalized momentum $\pi'_{\mu}$ is ill defined and cannot be easily made equal to the 
geometric momentum $p_{\mu}$. Nevertheless, if one keeps track of the factor 
$\sqrt{g_{\mu\nu}v^{\mu}v^{\nu}}$ when analyzing $S_{1}$ one can see that the corresponding 
Euler--Lagrange equations are the same as those for $S_{2}$ if one imposes the condition 
$g_{\mu\nu}v^{\mu}v^{\nu}=constant$. Thus, in this example $S_{1}$ and $S_{2}$ are equivalent 
as long as $g_{\mu\nu}v^{\mu}v^{\nu}$ is an integral of the motion. This condition is easily seen  
to be valid for $S_{2}$ due to the fact that  the Hamiltonian function for $L_{2}$ is equal to $L_{2}$.
Thus any solution related to $L_{2}$ will correspond to a solution for $L_{1}$ that satisfies, 
in this case a supplemental condition, $g_{\mu\nu}v^{\mu}v^{\nu}=constant$. 
Therefore, solutions for  $L_{1}$  that do not satisfy this ``physical'' condition will be un-physical 
since they will not be solutions related to  $L_{2}$. In this respect $L_{2} \subset L_{1}$.

The ``physical'' assumption $g_{\mu\nu}v^{\mu}v^{\nu}=constant$ is a key ingredient of the 
parallel transport considerations in  Einstein General Relativity (GR), which seems to be obeyed by nature.
After all, studying processes out there in the universe, especially these that are particularly far from our labs,
can be easily understood if this condition was satisfied and thus resulting in the corresponding 
geodesic Equation (\ref{geodesic equations}). However, the past few decades of studies on 
far away galaxies and the universe as a whole have brought some puzzling  results that have been
attributed to dark matter and dark energy phenomena that have not be confirmed in our local laboratories.
It is often commonly expected that dark matter and dark energy are probably a new kind of particles and/or fields 
that have not yet been experimentally discovered, but perhaps various upcoming efforts \cite{Di Valentino et al.(2020)} 
to address current discordances present between the different cosmological probes  could result in the detection 
of new interaction fields that may even be relevant to phenomenon of inflation. 
{Very recent research results, literally just submitted for publication by the authors, 
in effort to understand the Scale Invariant Vacuum (SIV) theory} \cite{Maeder79} 
{and its limitations has resulted in encouraging connections with standard models of inflation.}
Interestingly, these dark phenomena may be illuminated and could be understood quite well within 
the Integrable  Weyl geometry paradigm  that deviates from the standard Einstein GR 
parallel transport considerations \cite{MaedGueor19, MaedGueor20}. The Integrable Weyl geometry
does allow for departing from $g_{\mu\nu}v^{\mu}v^{\nu}=constant$. Thus, it provides a framework 
with a larger set of solutions, those solutions to $L_{1}$ that are not part of the $L_{2}$ space of solutions, 
to be explored for a better understanding of nature. 


In the above example, the mass multiplier $m$ in the Lagrangian was not relevant for the  
equations of motion (\ref{Euler-Lagrange equations}) but the mass was showing up as an
integral of the motion via the relevant energy-momentum dispersion relation $p_{\nu}p^{\nu}=m^2$.
This is usually the rest-mass of a particle. In general, however, when there are additional 
interactions, say electromagnetic, then the mass term in the Lagrangian is playing the role of
a coupling constant to the gravitational field but it can also manifest itself as a Lagrange multiplier.

Indeed, if one starts with the re-parametrization invariant Lagrangian
$L=qA_{\alpha}v^{\alpha}+m\sqrt{g_{\alpha\beta}(x)v^{\alpha}v^{\beta}}$,
which is usually interpreted as the relativistic 
Lagrangian of a massive particle due to the mass parameter $m$. 
The mass here is actually a coupling constant to the gravitational field $g_{\alpha\beta}$, 
just like the charge $q$  is playing the role of coupling constant to the 
electromagnetic vector potential $A_{\alpha}$.
By utilizing the reparametrization invariance and defining the
{proper time} $\tau$ such that: 
$d\tau=\sqrt{g_{\alpha\beta}dx^{\alpha}dx^{\beta}}\Rightarrow\sqrt{g_{\alpha\beta}v^{\alpha}v^{\beta}}=1$, 
then one can effectively consider $L=qA_{\alpha}v^{\alpha}+(m+\chi)\sqrt{g_{\alpha\beta}v^{\alpha}v^{\beta}}-\chi$
as our model Lagrangian. Here $\chi$ is a Lagrange multiplier to
enforce $\sqrt{g_{\alpha\beta}v^{\alpha}v^{\beta}}=1$ that breaks
the reparametrization invariance explicitly. Then one can write it
as $L=qA_{\alpha}v^{\alpha}+(m+\chi)\frac{g_{\alpha\beta}v^{\alpha}v^{\beta}}{\sqrt{g_{\alpha\beta}v^{\alpha}v^{\beta}}}-\chi$
and using $\sqrt{g_{\alpha\beta}v^{\alpha}v^{\beta}}=1$ one arrives
at $L=qA_{\alpha}v^{\alpha}+(m+\chi)g_{\alpha\beta}v^{\alpha}v^{\beta}-\chi$.
One can deduce a specific value for $\chi$ ($\chi=-m/2$) by requiring
that $L=qA_{\alpha}v^{\alpha}+m\sqrt{g_{\alpha\beta}(x)v^{\alpha}v^{\beta}}$
and $L=qA_{\alpha}v^{\alpha}+(m+\chi)g_{\alpha\beta}v^{\alpha}v^{\beta}-\chi$
produce the same Euler--Lagrange equations under the constraint $\sqrt{g_{\alpha\beta}v^{\alpha}v^{\beta}}=1$.
Then, by dropping the overall constant term, this finally results
in the familiar equivalent Lagrangian: $L=qA_{\alpha}v^{\alpha}+\frac{m}{2}g_{\alpha\beta}v^{\alpha}v^{\beta}$
where $\tau$ has the usual meaning of proper-time parametrization such that $\sqrt{g_{\alpha\beta}v^{\alpha}v^{\beta}}=1$.
This quadratic Lagrangian is often considered as more convenient to work with \cite{Pauli_1958} due to
a variety of unpleasant properties of the original re-parametrization invariant Lagrangian. 
In particular, Lagrangians quadratic in the velocity are preferred since the corresponding Hamiltonian
function is quadratic in the momentum and apparently is non-zero. 
However, if one is to keep track of all the manipulations above, then one would notice 
that the apparently quadratic Hamiltonian $H=\frac{m}{2}g_{\alpha\beta}v^{\alpha}v^{\beta} +\chi$
is identically zero when utilizing $g_{\alpha\beta}v^{\alpha}v^{\beta}=1$ and $\chi=-m/2$.
The next section is devoted to the justification of why the original re-parametrization invariant Lagrangian
may be more relevant for understanding the physical reality despite its unpleasant properties.
The importance of such re-parametrization invariant Lagrangian as a way of adding metric structure to 
an affine space has been emphasized previously by Randers in his paper 
``On an Asymmetrical Metric in the Four-Space of General Relativity''
along the various connections to the treatment of gravity and electromagnetism within a similar framework
\cite{Randers41}.

\section{Homogeneous Lagrangians \label{Homogeneous Lagrangians}}

Suppose one does not know classical physics, which is
mainly concerned with trajectories of point particles in some space
$M$ but is told that can derive it from a variational principle if
the right action integral $S=\int Ld\tau$ is used. By following the
above example, one would wonder: ``Should the smallest `time distance'
be the guiding principle?'' when constructing $L$. If yes, ``How
should it be defined for other field theory models?'' It seems that
a reparametrization-invariant theory can provide us with a metric-like
structure \cite{Rund_1966,Randers41}, and thus, a possible link between field
models and geometric models \cite{Rucker_1977}.

In the example of the relativistic particle (Section \ref{Relativistic Particle}
above), the Lagrangian and the trajectory parameterization have a
geometrical meaning. In general, however, parameterization of a trajectory
is quite arbitrary for any observer. If there is the smallest time
interval that sets space-time scale, then this would imply a discrete
space-time structure since there may not be any events in the smallest
time interval. The Planck scale is often considered to be such a special
scale \cite{Magueijo_and_Smolin}. Leaving aside recent hints for
quantum space-time from loop quantum gravity and other theories, one
should ask: ``Should there be any preferred trajectory parameterization
in a smooth 4D space-time?'' and ``Are we not free to choose the
standard of distance (time, using natural units $c=1$)?'' If so,
then {one should have a smooth continuous manifold and our
theory should not depend on the choice of parameterization}.

If one examines the Euler--Lagrange equations carefully: 
\begin{equation}
\frac{d}{d\tau}\left(\frac{\partial L}{\partial v^{\alpha}}\right)
=\frac{\partial L}{\partial x^{\alpha}},\label{Euler-Lagrange equations}
\end{equation}
one would notice that any homogeneous Lagrangian of order 
$n$ ($L(x,\alpha\vec{v})=\alpha^{n}L(x,\vec{v})$)
provides a reparametrization invariance of the equations 
under the transformations $\tau\rightarrow\tau/\alpha,\vec{v}\rightarrow\alpha\vec{v}$.
As a side remark, notice that for homogeneous Lagrangian in $x$, 
the Euler--Lagrange equations possess scale invariance upon rescaling of the coordinates $x$. 
In general, such symmetries are related to the freedom of choosing a system of units by
the laboratory observer. However, the symmetry is often broken due
to the natural scales relevant to the specific process under study.
Next, note that the action integral $S$ involves an integration that
is a natural structure for orientable manifolds ($M$) with an $n$-form
of the volume. Since a trajectory is a one-dimensional object, then
what one is looking at is an embedding:
\begin{equation}
\phi:\Bbb{R}^{1}\rightarrow M\label{trajectory embedding}.
\end{equation}
This means that the map $\phi$ pushes forward the tangential space
$\phi_{*}:T(\Bbb{R}^{1})=\Bbb{R}^{1}\rightarrow T(M)$, and pulls
back the cotangent space $\phi^{*}:T^{*}(\Bbb{R}^{1})=\Bbb{R}^{1}\leftarrow T^{*}(M)$.
Thus, a 1-form $\omega$ on $M$ that is in $T^{*}(M)$ ($\omega=A_{\mu}\left(x\right)dx^{\mu}$)
will be pulled back on $\Bbb{R}^{1}$ ($\phi^{*}(\omega)$) and there
it should be proportional to the volume form on $\Bbb{R}^{1}$ ($\phi^{*}(\omega)
=A_{\mu}\left(x\right)(dx^{\mu}/d\tau)d\tau\sim d\tau$),
allowing one to integrate $\int\phi^{*}(\omega)$: 
\[
\int\phi^{*}(\omega)=\int Ld\tau=\int A_{\mu}\left(x\right)v^{\mu}d\tau.
\]
Therefore, by selecting a 1-form $\omega=A_{\mu}\left(x\right)dx^{\mu}$
on $M$ and using $L=A_{\mu}\left(x\right)v^{\mu}$ one is actually
solving for the embedding $\phi:\Bbb{R}^{1}\rightarrow M$ using a
chart on $M$ with coordinates $x:M\rightarrow\Bbb{R}^{n}$.

The Lagrangian obtained this way is first-order homogeneous in the
velocity $v$ with very simple dynamics. The corresponding Euler--Lagrange
equation is $F_{\nu\mu}v^{\mu}=0$ where $F$ is a 2-form ($F=dA$);
in electrodynamics, this is the Faraday's tensor. If one relaxes the
assumption that $L$ is a pulled back 1-form and assume that it is
just a homogeneous Lagrangian of order one, then one finds a reparametrization-invariant
theory that has an important physics-related dynamics.

\subsection{Pros and Cons of Homogeneous Lagrangians of First Order\label{Pros and Cons}}

Although most of the features listed below are more or less self-evident,
it is important to compile a list of properties of the first-order
homogeneous Lagrangians in the velocity $\vec{v}$.\\

\noindent
Some of the good properties of a theory with a first-order homogeneous Lagrangian are: 
\begin{itemize}
\item[(1)] First of all, the action $S=\int L(x,\frac{dx}{d\tau})d\tau$ is a reparametrization invariant. 
Thus, there is no fictitious acceleration due to un-proper time parametrization (see Section \ref{Un-Proper Time});
\item[(2)] For any Lagrangian $L(t,x^{i},\frac{dx^{i}}{dt})$ one can construct
a reparametrization-invariant Lagrangian by enlarging the space from
${x^{i}:i=1,\dots,n}$ to an extended space-time ${x^{\mu}:\mu=0,1,\dots,n},x^{0}=t$
\cite{Goldstain_1980,2011AmJPh..79..882D}: 
$L(t,x^{i},\frac{dx^{i}}{dt})\rightarrow L(x^{\mu},\frac{dx^{\mu}}{dx^{0}})\frac{dx^{0}}{d\tau}$.
The Euler--Lagrange equations for these two Lagrangians are equivalent
as long as $v^{0}=dt/d\tau$ is well behaved and $\tau$ is also a reasonable
``time''-parametrization choice;
\item[(3)] Parameterization-independent path-integral quantization is possible
since the action $S$ is reparametrization invariant;
\item[(4)] The reparametrization invariance may help in dealing with singularities
\cite{Kleinert_1989};
\item[(5)] It is easily generalized to extended objects ($p$-branes) that is
the subject of Section \ref{Extended Objects}. 
\end{itemize}
The list of trouble-making properties in a theory with a first-order
homogeneous Lagrangian includes: 
\begin{itemize}
\item[(1)] There are constraints among the Euler--Lagrange equations \cite{Goldstain_1980}
since $\det\left(\frac{\partial^{2}L}{\partial v^{\alpha}\partial v^{\beta}}\right)=0$;
\item[(2)] It follows that the Legendre transformation ($T\left(M\right)\leftrightarrow T^{*}\left(M\right)$),
which exchanges velocity and momentum coordinates $(x,v)\leftrightarrow(x,p)$,
is problematic \cite{Gracia_and_Josep};
\item[(3)] There is a problem with the canonical quantization approach since
the Hamiltonian function is identically ZERO ($h\equiv0$) \cite{Nikitin-StringTheory}. 
\end{itemize}

\textls[-15]{Constraints among the equations of motion are not an insurmountable
problem since there are procedures for quantizing such theories \cite{Nikitin-StringTheory,Dirac_1958a,Teitelboim_1982,Henneaux_and_Teitelboim,Sundermeyer_1982}.
For example, instead of using $h\equiv0$ one can use some of the
constraint equations available, or a conserved quantity, as Hamiltonian
for the quantization procedure \cite{Nikitin-StringTheory}. Changing
coordinates $(x,v)\leftrightarrow(x,p)$ seems to be difficult but
it may be resolved in some special cases by using the assumption that
a gauge $\Lambda$ has been chosen so that $L\rightarrow L+\frac{d\Lambda}{d\tau}=\tilde{L}=const$.
The above-mentioned quantization difficulties would not be discussed since
they are outside of the scope of this paper. A new approach that turns
the problem $h\equiv0$ into a virtue and naturally leads to a Dirac-like
equation is under investigation and the subject of a forthcoming paper,
for some preliminary details see \cite{VGG_Varna_2002}. 
Currently the new quantization approach has resulted in interesting connections 
and new view points at some of the key properties of physical \mbox{systems   
\cite{2011AmJPh..79..882D,Gueorguiev2019}}.
Even though the connection to quantum physics is beyond the scope of the
current paper, we would like to point out that the results in \citep{2011AmJPh..79..882D,Gueorguiev2019}
concerned with the consistent quantization of re-parametrization invariance systems,
provide a new viewpoint on the choice of the Hamiltonian constraint and 
the meaning of process parametrization within a chosen quantization frame. 
In particular, it is shown that the positivity of the rest energy is  
related to the requirement of normalizability of the states by utilizing the 
Hamiltonian constraint into a quantum constrain ($\hat{H}\Psi=0$). In this respect,
the results in \citep{2011AmJPh..79..882D,Gueorguiev2019} are reproducing the familiar quantum results 
where the constraint eliminates the ghost states and the ordinary mass shell constraint is
related to the Klein--Gordon equation \cite{Todorov'78, Horwitz and Rohrlich'81}. }

\subsection{Canonical Form of the First-Order Homogeneous Lagrangians\label{Canonical Form}}

Hopefully, by now the reader is puzzled, and is wondering along the
following line of thinking: ``What is the general mathematical expression
for first-order homogeneous functions?'' In this section, the notion
of the {canonical form of the first-order homogeneous Lagrangian}
and why such a form may be a useful mathematical expression from a physics
point of view is justified.

First, note that any symmetric tensor of rank $n$ 
($S_{\alpha_{1}\alpha_{2}\dots\alpha_{n}}=S_{[\alpha_{1}\alpha_{2}\dots\alpha_{n}]}$,
where $[\alpha_{1}\alpha_{2}\dots\alpha_{n}]$ is an arbitrary permutation
of the indexes) defines a homogeneous function of order 
$n$ ($S_{n}(\vec{v},\dots,\vec{v})=S_{\alpha_{1}\alpha_{2}\dots\alpha_{n}}v^{\alpha_{1}}\dots.v^{\alpha_{n}}$)
in the velocity $v$. The symmetric tensor of rank two is denoted
by $g_{\alpha\beta}$. Using this notation, the {canonical
form of the first-order homogeneous Lagrangian} is defined as: 
\begin{eqnarray}
L\left(\vec{x},\vec{v}\right) & = & \sum_{n=1}^{\infty}\sqrt[n]{S_{n}\left(\vec{v},\ldots,\vec{v}\right)}=\label{canonical form}\\
 & = & A_{\alpha}v^{\alpha}+\sqrt{g_{\alpha\beta}v^{\alpha}v^{\beta}}+\ldots+\sqrt[m]{S_{m}\left(\vec{v},\ldots,\vec{v}\right)}.\nonumber 
\end{eqnarray}

Whatever is the Lagrangian for the matter, it should involve interaction
fields that couple with the velocity $\vec{v}$ to a scalar. Thus,
the matter Lagrangian $L_{matter}\left(\vec{x},\vec{v};Fields~\Psi\right)$
would depend also on the interaction fields. When the matter action
is combined with the action ($\int\mathcal{L}[\Psi]dV$) for the interaction
fields $\Psi$, then one obtains a full {background independent
theory}. Then, the corresponding Euler--Lagrange equations contain
``dynamical derivatives'' on the left-hand side and sources on the
right-hand side: 
\[
\partial_{\gamma}\left(\frac{\delta\mathcal{L}}{\delta(\partial_{\gamma}\Psi)}\right)=
\frac{\delta\mathcal{L}}{\delta\Psi}+\frac{\partial L_{matter}}{\partial\Psi}.
\]

There are many ways to write first-order homogeneous functions
\cite{Rund_1966}. For example, one can consider the following expression
$L\left(\vec{x},\vec{v}\right)=\left(h_{\alpha\beta}v^{\alpha}v^{\beta}\right)\left(g_{\alpha\beta}v^{\alpha}v^{\beta}\right)^{-1/2}$
where $h$ and $g$ are seemingly different symmetric tensors. However,
each one of these fields ($h$ and $g$) has the same source type
($\sim v^{\alpha}v^{\beta}$): 
\[
\frac{\partial L}{\partial h_{\alpha\beta}}=\frac{L\left(\vec{x},\vec{v}\right)}{h_{\gamma\rho}v^{\gamma}v^{\rho}}v^{\alpha}v^{\beta},\quad\frac{\partial L}{\partial g_{\alpha\beta}}=\frac{L\left(\vec{x},\vec{v}\right)}{g_{\gamma\rho}v^{\gamma}v^{\rho}}v^{\alpha}v^{\beta}.
\]
Theories with two metrics have been studied before \cite{Dirac_1979,Bekenstein_1993}.
However, at this stage of our discussion, it seems unclear why the
same source type should produce different \mbox{interaction fields}. 

Some other relevant examples come from the field of 
Finsler spacetime geometry and its applications to physics \cite{Goenner2008, Pfeifer2019}, 
in particular the work of Bogoslovsky seems to be culminating in an experimentally testable 
framework \cite{Bogoslovsky2007,Bogoslovsky2020}. In his approach, Bogoslovsky is utilizing 
a Lagrangian $ \sim f_0(\vec{n},\vec{v})^b \sqrt{1-v^2/c^2}$ that has an explicit isotropy breaking effect 
$\vec{n}$ due to a conformal factor $f_0(\vec{n},\vec{v})$ 
based on zeroth-order homogeneous function $f_0$ in the velocity of a particle $\vec{v}$.
The power $b$ is a parameter that at $b=0$ results in the usual ``gravity''-like interaction while when
$b=1$, then the system seems to be more involved in an ``electromagnetic''-like interaction.   
Using a standard gravity-like interaction along with a zeroth-order homogeneous function
is an alternative mathematical approach, $L = f_0(x,v) \sqrt{g_{\mu\nu}v^{\mu}v^{\nu}}$,
since one expects gravity to always be present. Furthermore, such functional form for $L(x,v)$
has a good justification based on the Ehlers--Pirani--Schild axiomatic approach to 
Finsler geometry along with some additional requirements on the functional form of $f_0$ \cite{Pfeifer2019}.
Unfortunately, besides the Bogoslovsky case where the isometry breaking field $\vec{n}$ and its consequences are
well studied, the approach based on $L = f_0(x,v) \sqrt{g_{\mu\nu}v^{\mu}v^{\nu}}$ 
is not yet helpful in understanding the general structure of $f_0$, 
the meaning of the interaction fields, which will be involved in its content, 
and their classification. Even more, it is not clear if electromagnetic phenomenon could be 
described appropriately with such particular Lagrangian.

The advantage of the canonical form of the first-order homogeneous
Lagrangian (\ref{canonical form}) is that each interaction field,
which is associated with a symmetric tensor, has a {unique matter source}
that is a monomial in the velocities: 
\begin{equation}
\frac{\partial L}{\partial S_{\alpha_{1}\alpha_{2}\dots\alpha_{n}}}
=\frac{1}{n}\left(S_{n}(\vec{v},\dots,\vec{v})\right)^{\frac{1-n}{n}}v^{\alpha_{1}}\dots.v^{\alpha_{n}}.\label{sources}
\end{equation}

Therefore, the canonical form (\ref{canonical form}) is a natural choice 
for further discussion of the first-order homogeneous Lagrangians. 
Moreover, if one embraces the 
{principle of one-to-one correspondence between an interaction field and its source},
then the canonical form of the first-order homogeneous Lagrangian 
(\ref{canonical form}) via (\ref{sources}) justifies, 
from the mathematical point of view, the presence of the electromagnetic
and gravitational fields in nature.
 
{If one could devise a unique procedure
to express any first-order homogeneous function in the canonical form
above by using only the first two terms, 
then this could be viewed as a mathematical explanation of the
unique physical reality of only two fundamental classical interactions
-- the electromagnetic and gravitational interactions.} 
The first suggestion for such a procedure is given in problem \ref{EMandGravityOnly}  
of the exercises in Section \ref{Exercises}. When applied to a Lagrangians that contain only 
electromagnetism and gravity only ($n=1\, \&\, 2$), then the procedure recovers the original Lagrangian.
For other Lagrangians it can produce effective electromagnetism and gravity-only Lagrangian.
However, the equivalence or the specific accuracy of the approximation to the original equations of motion 
is not yet clear.

Following the Randers and Finsler path one could formally split a general reparametrization invariant model based on
first-order homogeneous Lagrangian $L(x,v)$ into electromagnetic and gravitational-like interactions as an
even and odd part of $L(x,v)=L^{(-)}(x,v)+L^{(+)}(x,v)$. Where $L^{(\pm)}(x,v)=(L(x,v) \pm L(x,-v))/2$.
Electromagnetic effects are then related to $qA_{\mu}(x,\vec{v}/c):=\partial L^{(-)}(x,v)/\partial v^{\mu}$ where the {E\&M} 
four-vector
potential has velocity dependence only on the special 3D velocity $\vec{v}=d\vec{x}/dt$; thus, it is homogeneous function of order zero.
Example of such velocity dependent E\&M fields and its importance has been discussed by Carlip \cite{Carlip}. 
Next the gravitational effects could now be related to the even part of the Lagrangian by considering the corresponding 
Finslerian metric $m^2g_{\mu\nu}(x,v):=\partial^2 L^{(+)}(x,v)^2/\partial v^{\mu}\partial v^{\nu}$ that is expecting to 
be a homogeneous function of order zero in the velocity and thus to have dependence only on the special 3D velocity 
$\vec{v}=d\vec{x}/dt$ as well. Again the velocity dependent gravitational potentials are essential to the understanding 
of the relation to retarded potentials and so on \cite{Carlip}. 
{ This is a simple justification of why there are only gravity and
electromagnetism on a classical level where the test particles are probing the fields far from their sources.}
However, there are further details to be worked out, like the equations satisfied by these classical fields, and as to 
what level this procedure actually reproduces the original Lagrangian i.e. see exercise \ref{EMandGravityOnly}.
The answers to all these questions however depend on the specific Lagrangian to be studied.
An alternative is to study the canonical form (\ref{canonical form}) were the fields are velocity independent.

Thus, it is important to investigate the additional higher-order terms and their
relevance to our physical reality. However, before entering into such
a discussion, which is the focus of Section \ref{New Forces}, it
will be interesting to touch upon the geometric description behind
the {principle of reparametrization invariance.} 
In this respect, the next subsection will discuss examples of the relevant Lagrangians that 
provide an illustration of the power of that principle in the justification of important 
Lagrangian-based models such as relativistic point particle, strings, and $p$-branes 
in theoretical and mathematical physics.

\subsection{$E$-dimensional Extended Objects \label{Extended Objects}}

At the beginning of the current section, the classical mechanics of a point-like
particle has been discussed as a mathematical problem concerned with the embedding
$\phi:\Bbb{R}^{1}\rightarrow M$ (\ref{trajectory embedding}). The
map $\phi$ provides the description of the trajectory (the word line)
of the particle in the target space $M$. The actual coordinate realization
of the map $\phi$ depends on the choice of the Lagrangian $L$ and
the interaction fields in $M$ due to the other objects that are already
in $M.$ According to the canonical form of the first-order homogeneous
Lagrangians (\ref{canonical form}), a point particle would interact
with electromagnetic-like vector field $A_{\mu}(x)$ and gravitation-like
(symmetric rank 2 tensor) field $g_{\mu\nu}(x)$, as well as with
other possible classical long-range fields (\ref{sources}) that are
described via rank $n>2$ symmetric tensors $S_{\alpha_{1}\alpha_{2}\dots\alpha_{n}}(x)$.
 
These interaction fields can be viewed as an embedding of higher-dimensional
objects. For example, $A_{\mu}(x)$ may be viewed as an embedding of $M$
into space with the same dimension $m$, for electromagnetism, it
is $4D$ space into another $4D$ space. For gravity, it is about
$4D$ space into a $10D$ space since there are 10 independent entries
in a symmetric $4\times4$ matrix $g_{\mu\nu}(x).$ However, one does
not have to consider only the interaction fields for a point particle.
One can consider a more general extended object called $p$-brane.
In this sense, the classical mechanics of a point-like particle that
has been discussed as a problem concerned with the embedding $\phi:\Bbb{R}^{1}\rightarrow M$
is actually a $0$-brane that is a one-dimensional object. 
Although time is kept in mind as an extra dimension, 
one should not insist on any special structure associated with a time flow. 
For this reason there will be no $0$-label(s), which usually singles out 
the time component(s), in this section.

Let us think of an extended object as a manifold $E$ with dimension, denoted also
by $E,\dim E=E=p+1$ where $p=0,1,2,\dots$. In this respect, one has
to solve for $\phi:E\rightarrow M$ such that some action integral
is minimized. From this point of view, one is dealing with the mechanics
of a $p$-brane. In other words, how is this $E$-dimensional extended object,
which is representing the ``trajectory'' of a $p$-brane,
submerged in $M$, and what are the relevant interaction fields?
By using the coordinate charts on $M$ ($x:M\rightarrow\Bbb{R}^{m}$),
one can think of this as a field theory over the $E$-manifold with
a local fiber $\Bbb{R}^{m}$. Thus the field $\vec{\phi}$ is such that: 
\begin{equation}
\vec{\phi}:\:\phi^{\alpha}=x^{\alpha}\circ\phi:E\rightarrow M\rightarrow\Bbb{R}^{m}.
\label{embedding of extended object}
\end{equation}

Following the relativistic point particle discussion after equation (\ref{trajectory embedding}), 
but this time using the pullback $\phi^{*}$ of the embedding map $\phi$ in (\ref{embedding of extended object}),
one considers the space of the $E$-forms over the manifold $M$, 
denoted by $\Lambda^{E}\left(M\right)$ with dimension $D=\binom{m}{E}=\frac{m!}{E!(m-E)!}$. 
In a specific coordinate basis a general element 
$\boldsymbol{\Omega}$ in $\Lambda^{E}\left(M\right)$ has the form 
$\boldsymbol{\Omega}=\Omega_{\alpha_{1}\dots\alpha_{E}}dx^{\alpha_{1}}\wedge 
dx^{\alpha_{2}}\wedge \dots\wedge dx^{\alpha_{E}}$.
Since $m \ge E$ for such an embedding then there are $D$ 
linearly independent $E$-forms in  $\Lambda^{E}\left(M\right)$.
Let us use an arbitrary label $\Gamma=1,2,\ldots,D$ to index the different $E$-forms over $M$; 
thus, $\boldsymbol{\Omega}^{\Gamma}=\Omega_{\alpha_{1}\dots\alpha_{E}}^{\Gamma}dx^{\alpha_{1}}
\wedge dx^{\alpha_{2}}\wedge\dots\wedge dx^{\alpha_{E}}$ are total of $D$ linearly independent $E$-forms 
in  $\Lambda^{E}\left(M\right)$.

Next, let us introduce ``{{generalized velocity vectors}}''
with components $\boldsymbol\omega^{\Gamma}$ : 
\begin{eqnarray}
\boldsymbol\omega^{\Gamma} & = & \frac{\boldsymbol{\Omega}^{\Gamma}}{dz}=
\Omega_{\alpha_{1}\dots\alpha_{E}}^{\Gamma}\frac{\partial\left(x^{\alpha_{1}}x^{\alpha_{2}}\dots 
x^{\alpha_{E}}\right)}{\partial(z^{1}z^{2}\dots z^{E})},
\label{generalized velocity vectors}\\
dz & = & dz^{1}\wedge dz^{2}\wedge\dots \wedge dz^{E}.\nonumber
\end{eqnarray}
In the above expression (\ref{generalized velocity vectors}), 
$\frac{\partial\left(x^{\alpha_{1}}x^{\alpha_{2}}\dots x^{\alpha_{E}}\right)}{\partial(z^{1}z^{2}\dots z^{E})}$
stands for the Jacobian of the transformation from coordinates $\{x^{\alpha}\}$
over the manifold $M$ to coordinates $\{z^{a}\}$ over the embedded manifold $E$.
The Jacobians provide a natural basis for the corresponding space. Note that $\Gamma$ in the above expression 
is a place holder for a particularly interesting vector in that space or a specific coordinate.
For example, if these $\boldsymbol{\Omega}^{\Gamma}$ forms are taken to be the Jacobians, then we will use the short hand notation 
$Y^{\Gamma}=\frac{\partial\left(x^{\alpha_{1}}x^{\alpha_{2}}\dots x^{\alpha_{E}}\right)}{\partial(z^{1}z^{2}\dots z^{E})}$
with $\Gamma$ being the integer valued labeling function of the anti-symmetric ordered set, e.g., lexicographically, 
$\left\{ \{ \alpha_{1}, \alpha_{2} \dots \alpha_{E} \} : \alpha_{i} \in \{1,\dots , m \} \right\}$
corresponding to an element there; therefore, with values in the range $\{1, \dots, D \}$. 
Thus, while $\boldsymbol{\Omega}^{\Gamma}$ stands for a particularly interesting vector in that space $\Lambda^{E}\left(M\right)$,
its coordinates, in that space, then are $\Omega_{\alpha_{1}\dots\alpha_{E}}^{\Gamma}$.
The general $\boldsymbol{\Omega}^{\Gamma}$ becomes $Y^{\Gamma}$ 
when the coordinates are the corresponding Kronecker-delta functions:
$\Omega_{\alpha_{1}\dots\alpha_{E}}^{\Gamma}=\delta_{\Gamma,\boldsymbol{\alpha}}$.

In the case (\ref{embedding of extended object}) the pull-back of an $E$-form $\boldsymbol{\Omega}^{\Gamma}$ 
must be proportional to the volume  $dz=dz^{1}\wedge dz^{2}\wedge\dots \wedge dz^{E}$ over the manifold $E$, 
just as in the corresponding discussion after Equation (\ref{trajectory embedding}): 

\begin{eqnarray*}
\phi^{*}\left(\boldsymbol{\Omega}^{\Gamma}\right) & = & \boldsymbol\omega^{\Gamma}dz^{1}\wedge dz^{2}\wedge\dots \wedge dz^{E}=\\
 & = & \Omega_{\alpha_{1}\dots \alpha_{E}}^{\Gamma}\frac{\partial\left(x^{\alpha_{1}}x^{\alpha_{2}}\dots x^{\alpha_{E}}\right)}{\partial(z^{1}z^{2}\dots z^{E})}dz^{1}\wedge dz^{2}\wedge\dots \wedge dz^{E}.
\end{eqnarray*}
Therefore, it is suitable for integration over the $E$-manifold.
Thus, the action for the embedding $\phi$ is: 
\[
S\left[\phi\right]=\int_{E}L\left(\vec{\phi},{\boldsymbol{\boldsymbol\omega}}\right)dz
=\int_{E}\phi^{*}\left(\boldsymbol{\Omega}\right)=\int_{E}A_{\Gamma}(\vec{\phi})\boldsymbol\omega^{\Gamma}dz.
\]
This is a homogeneous function in $\boldsymbol\omega$ and is reparametrization
(diffeomorphism) invariant with respect to the diffeomorphisms of
the $E$-manifold. If one relaxes the linearity 
$L(\vec{\phi},\boldsymbol\omega)=\phi^{*}\left(\boldsymbol{\Omega}\right)=A_{\Gamma}(\vec{\phi})\boldsymbol\omega^{\Gamma}$
in $\boldsymbol\omega$, then the canonical expression for the first-order
homogeneous Lagrangian gives: 
\begin{eqnarray}
L\left(\vec{\phi},\boldsymbol\omega\right) & = & \sum_{n=1}^{\infty}\sqrt[n]{S_{n}\left(\boldsymbol\omega,\dots ,\boldsymbol\omega\right)}=\label{canonical p-brane L}\\
 & = & A_{\Gamma}\boldsymbol\omega^{\Gamma}+\sqrt{g_{\Gamma_{1}\Gamma_{2}}\boldsymbol\omega^{\Gamma_{1}}\boldsymbol\omega^{\Gamma_{2}}}+\dots \sqrt[m]{S_{m}\left(\boldsymbol\omega,\dots ,\boldsymbol\omega\right)}.\nonumber 
\end{eqnarray}

At this point, there is a strong analogy between the relativistic
point particle and a general $p$-brane. However, there is a difference
in the number of components. 
In particular, $\vec{x},\vec{v}$, and $\vec{\phi}=\vec{x}\circ\phi$ have the same number of components ($m=\dim(M)$), 
however, the ``generalized velocity'' 
$\boldsymbol\omega$ has a bigger number of components $D=\binom{m}{E}\geq m$
that are related to the Jacobians (\ref{generalized velocity vectors}) \cite{Fairlie_and_Ueno}.
The linearly independent general elements of this  space are labeled with the index $\Gamma$
to allow index contraction with a relevant interaction field $A_{\Gamma}$,
$g_{\Gamma_{1}\Gamma_{2}}$, or $S^{n}_{\Gamma_{1}\dots \Gamma_{n}}$.

Some specific examples of $p$-brane theories correspond to the following
familiar Lagrangians in theoretical and mathematical physics: 
\begin{itemize}

\item The Lagrangian for a 0-brane (relativistic point particle in an electromagnetic field, $\dim E=1$ 
and $\boldsymbol\omega^{\Gamma}\rightarrow v^{\alpha}=\frac{dx^{\alpha}}{d\tau}$) is: 

\begin{eqnarray*}
L\left(\vec{\phi},\boldsymbol\omega\right)=A_{\Gamma}\boldsymbol\omega^{\Gamma}+\sqrt{g_{\Gamma_{1}\Gamma_{2}}\boldsymbol\omega^{\Gamma_{1}}\boldsymbol\omega^{\Gamma_{2}}} & \rightarrow & L\left(\vec{x},\vec{v}\right)\\
L\left(\vec{x},\vec{v}\right)=qA_{\alpha}v^{\alpha}+m\sqrt{g_{\alpha\beta}v^{\alpha}v^{\beta}};
\end{eqnarray*}

\item The Lagrangian for a 1-brane (strings, $\dim E=2$) \cite{Nikitin-StringTheory} is: 
\[
L\left(x^{\alpha},\partial_{i}x^{\beta}\right)=\sqrt{Y^{\alpha\beta}Y_{\alpha\beta}},
\]
using the notation: 
\begin{eqnarray*}
\boldsymbol\omega^{\Gamma}\rightarrow Y^{\alpha\beta}=\frac{\partial(x^{\alpha},x^{\beta})}{\partial(\tau,\sigma)}=\det\left(\begin{array}{cc}
\partial_{\tau}x^{\alpha} & \partial_{\sigma}x^{\alpha}\\
\partial_{\tau}x^{\beta} & \partial_{\sigma}x^{\beta}
\end{array}\right)=\\
=\partial_{\tau}x^{\alpha}\partial_{\sigma}x^{\beta}-\partial_{\sigma}x^{\alpha}\partial_{\tau}x^{\beta}.
\end{eqnarray*}
In this case, the index $\Gamma$ for labeling the components of the
{{generalized velocity vector $\boldsymbol\omega^{\Gamma}$}} corresponds
to the set of pairs $\{\alpha,\beta\}$ out of $m$ elements. For
example, for $m=4$ this will be $4!/2!^{2}=6$ not $4$ like for
the standard velocity vector in \textbf{\textit{$M$}};

\item The Lagrangian for a general $p$-brane has the Dirac-Nambu-Goto term
(DNG) \cite{Pavsic_2001}: 
\[
L\left(x^{\alpha},\partial_{E}x^{\beta}\right)=\sqrt{Y^{\Gamma}Y_{\Gamma}}.
\]
\end{itemize}
Notice that most of the Lagrangians above, except for the relativistic
particle, are restricted only to gravity-like interactions. In the
case of the charged relativistic particle, the electromagnetic interaction
is very important. The corresponding interaction term for $p$-branes
is known as a Wess--Zumino term \cite{Bozhilov_2002}.

The above discussion can be viewed as a justification of important class
of model Lagrangian systems via the {principle of reparametrization
invariance} when applied to the mechanics of point particle as well
as to the mechanics of extended objects. The principle leads naturally to 
important, well-known, and studied electromagnetic-like ($n=1$) and gravity-like ($n=2$) 
interaction terms. However, the framework also suggests new possible fields ($n>2$). 
Thus, it is important to investigate the additional higher-order terms and their relevance
to our physical reality.

\section{Classical Forces Beyond Electromagnetism and Gravity \label{New Forces}}

So far, the focus of the paper has been to justify and encourage the study of models
based on first-order homogeneous Lagrangians by emphasizing their
general properties, their potential to provide a mathematical justification
of the observed macroscopic physical reality, and along the way, 
to set the stage for the study of diffeomorphism invariant mechanics
of extended objects by following the close analogy with the relativistic point particle. 
Consequently, it is important to understand these new terms in the canonical expression 
of the first-order homogeneous Lagrangians (\ref{canonical form}). 
In this respect, this section discusses the implications of such interaction terms 
beyond electromagnetism and gravity as given by (\ref{canonical form}). 

Before we go into more detail on this exploration,
it should be pointed out that the results which follow are somewhat similar,
in the sense that such behavior is unusual for a standard classical particle, 
to what has been observed in the a la Bogoslovsky studies. 
In particular, in the Bogoslovsky's model the mass dependent terms in the corresponding 
energy-momentum dispersion relation receive additional parameter $b$-dependence 
due to the presence of the specific conformal factor, in this respect 
one should probably talk, for example, about renormalization of the kinematical mass.
Another effect observed is the presence of zero momentum even when the velocity is zero
\cite{Goenner2008,Bogoslovsky2020}. 
The work of Bogoslovsky will not be considered in any further details since 
the focus of the current study is on a different mathematical structure of the 
first-order homogeneous Lagrangians than the one studied by Bogoslovsky.
In this respect the purpose of this section is to emphasize the unusual  behavior when
going beyond standard gravity and electromagnetism, that is pure $S_n$ type interactions with $n>2$ 
rather than confirming or comparing the details of such unusual effects.

Let us begin our journey by recognizing that one can circumvent the linear dependence,
$(\det(\frac{\partial^{2}L}{\partial v^{\alpha}\partial v^{\beta}})=0)$
due to the reparametrization symmetry, of the equations of motion
derived from $L=\sqrt[n]{S_{n}\left(\vec{v},\dots ,\vec{v}\right)}$
by adding an extra set of equations ($\frac{dL}{d\tau}=0$). This
way the equations of motion derived from $L=\sqrt[n]{S_{n}\left(\vec{v},\dots ,\vec{v}\right)}$
and $\frac{dL}{d\tau}=0$ are equivalent to the equations of motion
derived from $L=S_{n}\left(\vec{v},\dots ,\vec{v}\right)$. This is similar
to the discussion at the end of Section \ref{Relativistic Particle}.
As noticed before, this is a specific choice of parametrization such
that $g_{\alpha\beta}\left(x\right)v^{\alpha}v^{\beta}$ is constant.

If one focuses on a specific $n^{th}$-term of re-parametrization
invariant Lagrangian (\ref{canonical form}), that is, $L=\left(S_{n}\left(v\right)\right)^{1/n}$
in the parametrization gauge $S_{n}\left(v\right)=const$ then the
equations of motion are:
\begin{equation}
S_{n/\alpha/\beta}\frac{dv^{\beta}}{d\tau}=S_{n,\alpha}-S_{n/\alpha,\beta}v^{\beta}.\label{SnEQM}
\end{equation}

Here $S_{n,\alpha}$ denotes partial derivative with respect to $x^{\alpha}$
when $S_{n/\alpha}$ indicates partial derivative with respect to
$v^{\alpha}$. From this expression, it is clear that $n=2$ is a
model that results in velocity independent symmetric tensor $S_{n/\alpha/\beta}(v)$
that can be associated with the metric tensor. Usually, such a metric
tensor is assumed invertible and therefore the differential equations
can be written in the form acceleration as a function of velocity
and position. However, in general, the left-hand side $S_{n/\alpha/\beta}(v)$
goes as $v^{n-2}$ while the right-hand side as $v^{n}$ which will
result in the general behavior that the acceleration grows as $v^{2}$ at
most. This is consistent with the known velocity dependence of the equation of 
the geodesics as well as the equation of the geodesic deviations. 
The growth is not usually an issue since there is a limitation on the magnitude 
$v<c$ due to the finite propagation speed. Thus, for a suitably chosen units, $c=1$
one should have $|v^{\alpha}|\leq1$. However, if $n>2$ and if the maximum speed limit 1 
is reached along one coordinate, then there could be issues for keeping the system 
at rest with respect to another coordinate  direction since a term like 
$1/v^{n-2}$ will grow towards infinity when $v\rightarrow~0$.
For homogeneous, isotropic, and static $S_n$ background field, 
with invertible symmetric tensor $S_{n/\alpha/\beta}(v)$,
a particle described by \mbox{(\ref{SnEQM})} should be moving with a constant non-zero speed in all directions,
in order to avoid infinite acceleration effects.

\subsection{Simplest Pure $S_{n}(v)$ Lagrangian Systems\label{Simplest  Sn Lagrangian Systems}}

To further illustrate our point above and to gain a better understanding
of the $S_{n}(v)$ terms, let us consider the simplest possible pure $S_{n}(v)$
Lagrangian systems by assuming:
\begin{itemize}
\item Curvilinear coordinate system such that: $S_{n}(v)=f(t,r,w,u)$ where $w=dx^{0}/d\tau$
and $u=dr/d\tau$;
\item Static fields, that is: $S_{n}(v)=f(r,w,u)$;
\item Inertial coordinate system in the sense of Newtonian like space and
time separation, that is: $S_{t\dots tr\dots r}=0$ except for $S_{t\dots t}$
and $S_{r\dots r}$ components.
\end{itemize}
Thus, the expression for $S_{n}(v)$ takes upon the following form:
\begin{equation}
S_{n}(r,w,u)=\psi(r)w^{n}+\phi(r)u^{n}.\label{S_=00007Bn=00007D(r,w,u)}
\end{equation}
Notice that here the symbols $u$ and $w$ are used to denote the
spatial $r$ and time-like velocity coordinates instead of the previously
used symbol $v$. Later in the discussion, this symbol $v$ will be
used to denote the spatial speed in the Newtonian limit using coordinate-time
parametrization $v=dr/dt$. This way the corresponding equations
of motion for $L=S_{n}(r,w,u)$ are: 
\begin{eqnarray}
\frac{du}{d\tau} & = & -\frac{u^{2}\phi^{\prime}(r)}{(n-1)\phi(r)}
+\frac{1}{u^{n-2}}\frac{w^{n}\psi^{\prime}(r)}{n(n-1)\phi(r)},\label{u-equation}\\
\frac{dw}{d\tau} & = & -\frac{wu\psi^{\prime}(r)}{(n-1)\psi(r)}.\label{w-equation}
\end{eqnarray}

One can recognize the connection of the fields $\psi(r)$ and $\phi(r)$
to the energy and linear momentum of a particle by looking at the
generalized momentum: $p_{\alpha}=\frac{\partial L}{\partial v^{\alpha}}$.
In particular, $\psi(r)$ is related to the energy of the particle
$E=p_{0}=\frac{\partial L}{\partial w}$, especially when considering
$\tau=x^{0}=ct$ in {co-moving frame} $u/w=v/c\approx0$
using coordinate-time parametrization where $w=1$. In this respect,
if the energy of the particle is conserved then $\psi(r)=constant$
and therefore $\psi^{\prime}(r)=0$. The ``radial'' acceleration
at macroscopic scales is then: 
\begin{equation}
a_{r}=\frac{dv}{dt}=-\frac{v^{2}\phi^{\prime}(r)}{(n-1)\phi(r)}.
\end{equation}
Here, the speed of light $c$, the maximum speed of propagation cancels
out and $u=dr/d(ct)=v/c$ is related to the spatial speed $v$. If
$n=2$ and $\phi(r)=br$ then one recovers the usual kinematical expression
for the normal acceleration of a particle moving in a circular orbit
($a_{n}=v^{2}/r$).

In general, however, depending on the specifics of the model and details
of $\phi(r)$ one may obtain deviations from the flat space or the
metric model for gravity. Such new models and forces could be relevant
at large/cosmological scales where the dark-matter problem manifests
itself in the deviation of the kinematical acceleration from the anticipated
gravitational acceleration in galaxies and clusters of galaxies. Depending
on the sign of $\phi^{\prime}(r)/\phi(r)$ this term could be ``dissipative'' 
in the sense that the system will settle at $v=0$ after a sufficiently long time
if the sign of $\phi^{\prime}(r)/\phi(r)$ is positive. If the sign
is negative then one has ``repulsive gravity'' that could be relevant
to the dark-energy problem since the system will have an unstable $v=0$ configuration. 
Any further speculations about this equation
are poorly justified without any underlining theory that predicts
$\phi(r)$ and compares it to experimental observations.

At the microscopic scale, however, one may have $\psi^{\prime}(r)\neq0$.
This could suggest that the energy $p_{0}$ may not be conserved due
to the interaction of the particle with the environment; thus, it
may be subject to energy exchange. However, $p_{0}$ should nevertheless
be conserved since the model under consideration has no explicit coordinate-time
dependence. This can be illustrated using the equation for $w$ (\ref{w-equation}).
The equation can be rewritten in a form that makes it easy to be integrated
and to see the conservation of the energy $p_{0}$: 
\begin{eqnarray}
d\ln(w) & = & -\frac{1}{(n-1)}d\ln(\psi)\Rightarrow\nonumber \\
\psi(r)w^{n-1} & = & constant=p_{0}/n\label{p0-conservation}
\end{eqnarray}
The corresponding generalized linear momentum then will be:
\begin{eqnarray}
p_{r}=n\phi(r)u^{n-1}=
\frac{\phi(r)}{\psi(r)}p_{0}\left(\frac{v}{c}\right)^{n-1}\label{p-momentum}
\end{eqnarray}
This shows that the model considered does not have the Bogoslovsky
non-zero rest momentum behavior and the effects due to $S_n$ fields 
on the velocity-momentum relation seems to be non-linear and highly ``relativistic''. 

Looking back at the first Equation (\ref{u-equation}) for $u,$ when
$n=2$ the spatial force has two parts, one is velocity independent
force proportional to $\propto\frac{p_{0}^{2}\psi^{\prime}(r)}{2^{3}\psi(r)^{2}\phi(r)}$,
and the other part could be ``dissipative'' if the sign of $\phi^{\prime}(r)/\phi(r)$
is positive or ``repulsive gravity'' if the sign of $\phi^{\prime}(r)/\phi(r)$
is negative as discussed earlier.

For $n>2$ the physics interpretation of the equation of motion (\ref{u-equation})
leads to unusual behavior: 
\begin{equation}
\frac{(n-1)}{w^{2}}\frac{du}{d\tau}=-\frac{v^{2}\phi^{\prime}(r)}{c^{2}\phi(r)}
+\frac{c^{n-2}}{v^{n-2}}\frac{\psi^{\prime}(r)}{n\phi(r)}.\label{unusual behavior}
\end{equation}

It seems that an observer cannot detect a particle to be in complete
rest ($u/w=v/c=0$) for a finite time interval $\Delta t$. If the
speed of a particle was zero ($v=0$) at some moment then the particle
should have an infinite acceleration $du/d\tau$ at that moment since
$du/d\tau\propto w^{2}v^{2-n}\rightarrow\infty$. Thus, the particle
will instantaneously leave the state $v=0$ for a non-zero velocity
state rather than staying in the zero velocity state. Depending on
the details of the fields $\psi^{\prime}(r)$ and $\phi^{\prime}(r)$
there may not be a zero external force configuration for such  a particle
in general. Nevertheless, specific fields $\psi^{\prime}(r)$ and
$\phi^{\prime}(r)$ may allow for zero acceleration state $du/d\tau=0$
and non-zero spatial velocity state $u/w=v/c\neq0$: 
\[
v^{n}=\frac{c^{n}\psi^{\prime}(r)}{n\phi^{\prime}(r)}.
\]
However, such state would imply that the observer cannot be in the
co-moving frame of the particle anymore since the coordinate time $t$
will not be synchronized with the ``proper-time'' $\tau$ of the
particle $dw/d\tau\neq0$ and thus $w=dx^{0}/d\tau\neq constant$.

The above-discussed pathology is strikingly similar to the manner in which
quantum mechanical particles behave: Particles cannot be localized
with speed as close to zero as one wishes to; even more, the conservation
of energy needs to be amended due to external fields (\ref{p0-conservation}).
Therefore, such terms with $n>2$ may play important role in the understanding
of the mechanism behind the inflation driven early stage of the universe,
as well as in the derivation of the Dirac equation containing fundamental
sub-atomic interactions beyond electromagnetism and gravity (for preliminary 
discussion see \cite{VGG_Varna_2002,VGG_Cincinnati_2003}).
It should be noted that such pathological behavior 
may be attributed to radiation-reaction or the problem of self-force 
(see i.e. Abraham–Lorentz and  Abraham–Lorentz–Dirac forces \cite{Rohrlich00,Dirac38})
which is resolved in quantum electrodynamics via renormalization that requires 
adding higher order counter-terms to the Lagrangian \cite{Polchinski84}.

Not being able to observe a particle at rest seems somewhat in contradiction
to our classical physics reality. However, the more appropriate Lagrangian
should take into account that ``empty space'' has Minkowski geometry:
\[
L=m\sqrt{\eta_{\alpha\beta}v^{\alpha}v^{\beta}}+\kappa\sqrt[n]{S_{n}\left(\vec{v},\ldots,\vec{v}\right)}.
\]
Here $\eta_{\alpha\beta}=(1,-1,\ldots,-1)$ is the Lorentz invariant
metric tensor. For Lagrangians that contain gravity ($S_{2}(v)$ term)
the problem for spatial velocity limit $v\rightarrow0$ does not exist
as discussed above for the case $n=2$. In the non-relativistic limit
($v/c\rightarrow0$), the present model of pure $S_{n}$ interaction
in Minkowski spacetime results in an acceleration $\frac{dv}{d\tau}$
that is the same up to $O(v^{2})$ terms for $L=const$ parametrization
as well as for $\sqrt{\eta_{\alpha\beta}v^{\alpha}v^{\beta}}=const$
parametrization. Thus the non-relativistic limit cannot distinguish
these two choices of parametrization.

\subsection{Choice of Proper Time Parametrization\label{Proper Time}}

It was mentioned earlier that for parameter independent homogeneous Lagrangians
of order $\alpha$, one has $h=(\alpha-1)L$ and thus $dL/d\lambda=0$
except for $\alpha=1$ that singles out first-order homogeneous Lagrangians.
When working with re-parametrization invariant Lagrangian model, one can
choose parametrization so that $Ld\lambda=d\tau$ or effectively thinking of $L(x,v)=const$.
This brings the homogeneous Lagrangians of first order back in the
family $dL/d\lambda=0$. 

This appears to be the choice of parametrization
to be made $\lambda \rightarrow \tau$ if the structure of $L$ is not known.
However, it seems that $\sqrt{g_{\alpha\beta}v^{\alpha}v^{\beta}}=const$
is preferred \cite{Randers41} as physically more relevant due to its connection to
the lifetime of unstable elementary particles. Especially, due to
the lack of experimental evidence that the lifetime of charged elementary
particles is affected by the presence of electromagnetic fields. This
can be related to the observation that for any Lagrangian of the form
$L=v^{\mu}A_{\mu}(x,v),$ where $x$ is space-time coordinate and
$v$ is a world-line velocity vector (4-vector for 3+1 space-time),
one can define a velocity dependent symmetric tensor
$g_{\alpha\beta}\left(x,v\right)=\frac{1}{2}(A_{\alpha/\beta}\left(x,v\right)+A_{\beta/\alpha}\left(x,v\right))$
where $A_{\beta/\alpha}\left(x,v\right)$ denotes partial derivative
with respect to $v^{\alpha}$ of $A_{\beta}\left(x,v\right)$. Then
one can show that $\frac{d}{d\lambda}\left(v^{\alpha}g_{\alpha\beta}\left(x,v\right)v^{\beta}\right)=0$
along the trajectory determined by the Euler--Lagrange equation for
$L=v^{\mu}A_{\mu}(x,v)$ -- just like the usual geodesic equation of
motion as in the discussion presented in Section \ref{Relativistic Particle}.
This symmetric tensor $g_{\alpha\beta}\left(x,v\right)$ does not depend on
the velocity independent electromagnetic vector potential $A_{\mu}(x)$
and thus the length of the vector as calculated with $g_{\alpha\beta}\left(x,v\right)$
is not affected by the presence of electromagnetic interaction. Therefore,
{a proper time} parametrization that coincides with the traditional
definition: $d\tau=\sqrt{g_{\alpha\beta}(x,v)dx^{\alpha}dx^{\beta}}$
can be introduced. 

The name of this special choice of  $\tau$ parametrization
derives from the fact that it is generally covariant and thus independent
of the observer's coordinate system and can be interpreted as the passing of time
measured in the rest frame of the system under study. Therefore, it
is often of the form $d\tau=\sqrt{g_{00}(t)}dt$ and thus can be integrated
along the laboratory coordinate time $t$. The laboratory coordinate
time $t$ is up to the observer at rest as part of the laboratory measuring
tools for various processes. Unfortunately, for first-order homogeneous
Lagrangians, one has $v^{\alpha}g_{\alpha\beta}\left(x,v\right)v^{\beta}=0$
because $A_{\mu}(x,v)$ is a homogeneous function of zero degree and
thus $v^{\beta}A_{\mu/\beta}\left(x,v\right)=0$. This seems to make
it difficult to define the {proper time} parametrization in
the usual way: $d\tau=\sqrt{g_{\alpha\beta}(x,v)dx^{\alpha}dx^{\beta}}$
for such first-order homogeneous Lagrangians $L=v^{\mu}A_{\mu}(x,v)$.

In this respect, for first-order homogeneous Lagrangians in the velocity,
it is not clear if one has to choose {``proper time''} parametrization
so that $L=const,$ or $\sqrt{g_{\alpha\beta}v^{\alpha}v^{\beta}}=const,$
or $L-A_{\mu}(x)v^{\mu}=const$. The choice $L-A_{\mu}(x)v^{\mu}=const$
may very well be a suitable choice since the weak and the strong
forces do have an effect on the lifetime of elementary particles;
for example, neutrons are unstable in free space but stable within nuclei. 
In connection to this, note that the other terms beyond gravity
($S_{n}$ with $n>2$) are seemingly related to the internal degrees
of freedom of the elementary particles. This should become more clear
once a non-commutative quantization ($v\rightarrow\gamma$) is applied
to the re-parametrization invariant Lagrangian, which will be discussed
elsewhere (for some preliminary results see \cite{VGG_Varna_2002,VGG_Cincinnati_2003}). 
Unfortunately, it is not clear how to extract the $A_{\mu}(x)v^{\mu}$ component 
of any first-order homogeneous Lagrangian $L$ mathematically, 
which is applicable to a physically relevant process,
that is not assuming electromagnetic interaction a priory.
Mathematically, one can extract $A_{\mu}(x)$ from first-order homogeneous
Lagrangian $L$ by considering $A_{\mu}(x)=L(x,v)_{/\mu}=p_{\mu}$ at $v^\alpha \rightarrow 0$; 
however, physically $v^0$ should never be zero.
Nevertheless, in the case of the Simplest Pure $S_{n}(v)$ Lagrangian Systems discussed
in the previous subsection, one can define {``proper time''}
parametrization under certain conditions. 

Based on the general discussion after Equation (\ref{SnEQM}) and 
on the specific example Equation ($\ref{u-equation}$), 
the condition for reasonable parametrization such as {``proper time''}
for $S_{n}(v)$ is surprisingly restrictive.  It demands $n=2$ so that the laboratory
clock could be at rest with respect to the particle studied. If $n>2$
there is this pathological behavior that moves the particle ``instantaneously'' away
from the rest frame of the clock. Thus, only $n=2$ allows for a rest
frame within the model Lagrangians discussed. 
Then by using the conservation of $p_{0}$ (\ref{p0-conservation})
one has: $2\psi(r)cdt=p_{0}d\tau$ where almost everything is a constant ($c$ and $p_0$)
and $\psi(r)$ seems to be related to the gravitational potential
at the location $r$ where the particle is. In conclusion, it seems
that {``proper time''} parametrization is only possible for
$n=2$ systems based on the analyses of the Simplest Pure $S_{n}(v)$
Lagrangian Systems and the discussion above. Thus, gravity is essential
for the notion of the {``proper time''} parametrization and
no other Simple $S_{n}(v)$ Lagrangian System provides an alternative
parametrization that makes sense as the passing of time in the rest frame
of a particle.

To conclude this section, one may naively extrapolate the scale at
which such new forces may be dominant. Considering that electromagnetic
forces are relevant at an atomic and molecular scale when gravity is
dominating the solar system and at galactic and cosmological scales,
then one may deduce that terms beyond gravity may be relevant at galactic
and intergalactic scales. Along this line of reasoning, a possible
determination of the structure of such forces from the velocity distribution
of stars in galaxies is an interesting possibility. In this respect,
such forces can be of relevance to the dark matter and dark energy
cosmology problems. The pathological $dv/d\tau\rightarrow\infty$
when $v\rightarrow0$, behavior of pure $S_{n}$ for $n>2$ interactions
could also be of relevance to inflation models. Finally, as already
mentioned, such terms are essential for bringing in fields beyond the electromagnetic fields 
into the Dirac equation when considering the quantization of the first-order homogeneous
Lagrangians in the velocity.

\subsection{Fictitious Accelerations in Un-proper Time Parametrization\label{Un-Proper Time}}
In the previous section, we discussed the concept of proper time parametrization.
The coordinate time is evidently another choice of  time parametrization.
In general, if the action is reparametrization invariant then one should be able 
to use any choice of time parametrization for a process. However, if the 
action is not  reparametrization invariant then one may find a puzzling phenomenon
due to the choice of time parametrization for a process. The presence of a
fictitious acceleration in un-proper time parametrization of 
non-reparametrization invariant action is the topic of the current section.

In the framework of Special and General Relativity, Carlip derived  
that retardation effects resulting from velocity dependent field potentials  
give rise to forces that are linear (for electromagnetism) and quadratic (for gravity)
extrapolations pointing towards the instantaneous source location \cite{Carlip}. 
Note that this remarkable result is about the force on a test particle that 
does not alter (by definition) the overall field produced by the source.

The superposition of fields is a hallmark of the Maxwell equations for electromagnetism,
however, this is not the case for gravity. Therefore, the overall gravitational field of 
a large realistic gravitationally bound system is somewhat more complicated. However, for a realistic two body system, 
the superposition of two or more fields does not represent the correct gravitational field needed to assess the motion of the bodies.
Within the classical Lagrangian approach, the simplest option that may show some superposition-like properties could 
be based on the familiar quadratic Lagrangian $ L_2(x,v)=g_{\mu\nu}v^\mu v^\nu$. 
For a point particle at point $x_0^{\mu}$ in the space-time such Lagrangian can clearly exhibit superposition 
up to a leading order term due to $N$ sources that are far from each other so that:
\begin{eqnarray}
g^{\mu\nu}=\eta^{\mu\nu}+\sum^N_{i=1} {g_i^{\mu\nu}(x_0;{x_i,v_i})},\quad
g_i^{\mu\nu}(x_0;{x_i,v_i})=\frac{2Gm_i}{(r(v_i))^3}\sigma_i^\mu \sigma_i^\nu.             \label{g-superposition}
\end{eqnarray}

In the above equation we follow Carlip's notation where $r(v_i)$ is the velocity dependent distance that is covariant. 
The subscripts $i>0$ is reserved for the point sources, 
while $i=0$ or no subscript is reserved for the test particle location. 
Note that no superscripts for $x_i$ and $v_i$ indicate the position and velocity four-vectors in 4D space-time.
Finally, due to the linearity of the Euler--Lagrange equations of motion derived from the action $A_2=\int  L_2(x,v) dt$ 
with respect to the corresponding Lagrangian it follows that there will be a superposition of the gravitational fields.
Notice that the coordinate time of the test particle $x_0^0=c t$ is naturally the time parameter to describe 
the evolution within the action integral. The above setup has a potential to agree with Carlip's derivations in \cite{Carlip}.

Now, let us consider the case when there are two bodies with a significant deviation from the test-particle idea.
In this case, one may want to start with the construction above for the metric field,
since it may stand the chance of almost linear superposition when the bodies are sufficiently far apart. 
However, in the setting up of the action $A=\int  L(x,v) d\tau$ one has to decide on 
what would be the meaning of the time-like parameter $\tau$. For example, in the case of the Planet-Moon system, 
should the time be the Planet time or the Moon time or the center of mass time? And if we choose the center of mass,
would the center of mass point-like potential be a good enough approximation of the true gravitational field?
In a true general relativistic approach the choice of coordinate system is irrelevant but this path is long and difficult to walk.
Instead, let us consider an equivalent Lagrangian formulation that has reparametrization invariant action: 
$A_1=\int  L_1(x,v) d\tau=\int \sqrt{ L_2(x,v)} d\tau$. 
Note that now the linear superposition of gravitational fields is most likely violated. 
Since this is parametrization independent, we can consider $\tau$ to be any coordinate time we desire.
Such expression for the gravitational part of the Lagrangian is a standard choice in Special and General Relativity along with 
the condition $g_{\mu\nu}v^\mu v^\nu=\pm1$, where the choice $\pm$ depends on the choice of the metric signature.
This choice means that the proper-time $\tau$ of the test particle has been chosen as the overall time parameter.
As a result one has $\dot{ L}=0$ along the path of the test particle and the equations of motion  derived from $A_2$ or $A_1$ 
are the same. What happens if one relaxes this choice ($\dot{ L}=0$)?

If one derives the Euler-Lagrange equations, one would find that the equations derived from the action: 
$A_f=\int f( L(x,v)) dt$ are:
\begin{eqnarray}
\frac{dp_\mu}{dt}=\partial_\mu  L-\frac{f''}{f'} \dot{ L}p_\mu,
\; p_\mu =\frac{\partial  L}{\partial v^\mu}\quad
\Rightarrow\; \frac{dp_\mu}{dt}=\partial_\mu  L
+\frac{\dot{ L}}{2 L}p_\mu,\; 
\text{when}\; f( L)=\sqrt{ L}.                                       
\label{fictitious-force}
\end{eqnarray}
From the above expressions one can see that if the time parameter is such that $\dot{ L}=0$ then the extra term in the right-hand-side  
will vanish  for any reasonable action $A_f$. Thus, all such  Euler--Lagrange equations will be equivalent. While this is highly desirable, 
one is very likely to use unsuitable time parameter and/or metric field due to deviation of nature from our ideal abstract model and
their inability to account for everything. Thus, $\dot{ L}\ne 0$ is a very likely situation resulting in a fictitious force.

For deviations from Einstein GR based on the Integrable Weyl geometry, 
in particular the Scale Invariant Vacuum (SIV) theory \cite{Maeder79}, 
one often considers a metric $g^{\mu\nu}=\lambda^{-2}g_{\rm GR}^{\mu\nu}$ along with a metric connection 
$\kappa_\mu=-\partial_\mu \ln{\lambda}$.
In SIV, one has $\lambda(t)\propto t^{-1}$ (only time dependence) and in the weak field limit of homogenous and isotropic space 
one has usually an extra acceleration: $\kappa_0 \vec{v}=\vec{v}/t$, where $t$ is the cosmic time since the Big Bang \cite{Maeder20}.
If we adopt the same view about our model metric $g^{\mu\nu}$ in $ L$, then we have:
\begin{eqnarray}
\frac{dv_\mu}{dt}=\partial_\mu  L+\kappa_0 v_\mu,\; \text{when}\; 
 L=g_{\mu\nu}v^\mu v^\nu.                                                
\label{fictitious-acceleration}
\end{eqnarray}

Thus, one can see the appearance of fictitious force that enhances the motion of a particle that acts as a non-conservative force.
Such  fictitious force seems to be reflecting the deviation of our models from the true reality. There could be a lot of things coming into
interaction with our gravitating system as well as internal changes such as tidal effects that transfer rotational energy into internal energy and so on.
In this respect the force is dissipative-like, since mechanical energy is transferred into  internal energy of the system. 
This should not affect the overall gravitational field but our inability of utilizing the true proper time parametrization 
for the description of the system will bring in an apparent violation of the usual conservation laws. Thus, the use of 
un-proper ({un-proper seems better name than improper, since any time parametrization should be ok, 
but only in proper time parametrization one will see clearly the conservation of familiar quantities})
time parametrization results in extra fictitious terms.

To understand better the effect consider the original generalized momentum case (\ref{fictitious-force}) and 
$ L=\lambda^{-2}  L_{\rm GR}$ with $d( L_{\rm GR})/dt=0$, which results in:
\begin{eqnarray}
\frac{dp_\mu}{dt}=\partial_\mu  L+\kappa_0 p_\mu,\quad  \text{where}\; 
\kappa_0=-\dot{\lambda}/{\lambda}.                                                        
\label{fictitious-force}
\end{eqnarray}

If we utilize the usual expectation for our model Lagrangian $  L$  that leads to the usual
energy and momentum conservation  ($\partial_t  L=0$ and $\partial_\phi  L=0$), that is,
absence of explicit time and angle $\phi$ dependence  for our model Lagrangian $  L$.
Then the corresponding energy, given by the quadratic Hamiltonian $(h\propto p_\mu v^\mu-L=g_{\mu\nu}v^\mu v^\nu)$  
and angular momentum $(J)$ conservation equations are modified and 
result in the following new expressions:
\begin{eqnarray}
\frac{dh}{dt}=\frac{d (\lambda^{-2}  L_{\rm GR})}{dt}= 2\kappa_0 h\quad
\Rightarrow \quad\frac{\dot{h}}{h}=2\kappa_0,                                                      
\label{mechanic-energy}\\
\frac{d J}{dt}=\partial_\phi  L+\kappa_0  J=\kappa_0  J\quad
\Rightarrow\quad\frac{\dot{ J}}{ J}=\kappa_0.                                              
\label{angular momentum}
\end{eqnarray}

The practical meaning of the new results is the possibility to observe non-conserva\-tion effects 
$\dot{h}\ne0$ and $\dot{J}\ne0$ 
when our experiment reaches accuracy resulting in fractional uncertainty compatible to 
$\kappa_0 \delta{t}$ at high time resolution. 
Even though the expressions above are written as derivatives these are actually very small 
effects that accumulate over an extended period of observational data. 
These non-conservation effects are usually buried in much bigger fluctuations of 
the corresponding quantities as seen in actual astronomical observations
that are usually explained by tidal effects and similar dissipative processes. 
{Recent research results suggest possible new viewpoint for understanding and probably explaining puzzling 
measurement results within the Solar System -- the paper is in preparation by the authors and in collaboration with  
Prof. M. Krizek.}
The main point here is the presence of such effects due to un-proper time parametrization 
of a process along with non-reparametriosation invariant action for that process.

\section{The Background Fields and Their Lagrangians \label{Field Lagrangians}}

The uniqueness of the interaction fields and their source types has
been essential for the selection of the matter Lagrangian (\ref{canonical p-brane L}).
The first two terms in the Lagrangian are easily identified as electromagnetic
and gravitational interaction. The other terms describe new classical
forces. It is not yet clear if these new terms are actually present
in nature or not, so one shall not engage them actively in the following
discussion but our aim is to start preparing the stage for such research
and discussions. At this point, one has a theory with background fields
since the equations for the interaction fields are not known. To complete
the theory, one needs to introduce actions for these interaction fields.

One way to write the action integrals for the interaction fields $S_{n}$
in (\ref{canonical p-brane L}) follows the case of the $p$-brane
discussion. There, one has been solving for $\phi:E\rightarrow M$
by selecting a Lagrangian that is more than a pull-back of an $E$-form
over the manifold $M$. In a similar way, one may view $S_{n}$ as
an $M$-brane field theory, where $S_{n}:M\rightarrow S_{n}M$ and
$S_{n}M$ is the fiber of symmetric tensors of rank $n$ over $M$.
This approach, however, cannot terminate itself since new interaction
fields would be generated as in the case of $\phi:E\rightarrow M$.

Another way assumes that $A_{\Gamma}$ is an $n$-form. Thus, one
may use the structure of the external algebra  $\Lambda\left(T^{*}M\right)$
over $M$ to construct objects proportional to the volume form over
$M$. For any $n$-form $(A)$ objects proportional to the volume
form $\Omega_{{\rm Vol}}$ can be constructed by using operations
in $\Lambda\left(T^{*}M\right)$, such as the external derivative
$d$, external multiplication $\wedge$, and the Hodge dual $*$.
For example, $A\wedge*A$ and $dA\wedge*dA$ are forms proportional
to the volume form $\Omega_{{\rm Vol}}$.

{The next important ingredient comes from the symmetry in the
matter equation}. That is, if there is a transformation $A\rightarrow A^{\prime}$
that leaves the matter equations unchanged, then there is no way to
distinguish $A$ and $A^{\prime}$ by experiments and measurements
via the matter that is obeying these equations. Thus the action for
the field $A$ should obey the same symmetry (gauge symmetry) as 
those found in the equations of motion for the matter.

\subsection{Justifying the Electromagnetic Action \label{E&M FF action}}

Let us consider now the matter equation for 4D electromagnetic interaction
which is $d\vec{v}/d\tau=F\cdot\vec{v}$ where $F$ is the 2-form obtained
by differentiation of the 1-form $A$ ($F=dA$), and the gauge symmetry
for $A$ is $A\rightarrow A^{\prime}=A+df$ since the external differential
operator $d$ obeys $d^{2}=0$. The reasonable terms, which
can result in the volume form $\Omega_{{\rm Vol}}$ for the field
Lagrangian $\mathcal{L}(A)$ of a 1-form field $A$, are then: $A\wedge*A,$
$dA\wedge dA$, and $dA\wedge*dA$ and of course $A\wedge A\wedge A\wedge A$.
The first and last terms do not conform with the gauge symmetry $A\rightarrow A^{\prime}=A+df$
and the second term $(dA\wedge dA)$ is a boundary term since $dA\wedge dA=d\left(A\wedge dA\right)$
that gives $\int_{M}d\left(A\wedge dA\right)=A\wedge dA$ at the boundary
of $M$; this term is interesting in the quantum Hall effect. Therefore,
one is left with a unique action for fields based on a one-form $A=A_{\mu}(x)dx^{\mu}$
that respects the gauge symmetry of the corresponding Euler--Lagrange
equations of motion for matter: $A\rightarrow A^{\prime}=A+df$ --
this is exactly the electromagnetic field generated by moving charges 
$j^\mu=\rho v^\mu$  and  described by the \mbox{standard action}: 
\[
S\left[A\right]=\int_{M}dA\wedge*dA+A_\mu j^\mu=\int_{M}F\wedge*F+A_\mu j^\mu.
\]
Note that if $F$ was considered as a fundamental field rather than $A$ then in 4D 
one can also consider the term $F\wedge F$. However, as soon as one recognizes that 
$F=dA$ then this becomes the boundary term ($dA\wedge dA$) discussed above. 
Furthermore, once $F=dA$ is recognized as a two-form and expressed in
the coordinate basis $F_{\mu\nu}dx^{\mu}\wedge dx^{\nu}$ then one
can also consider a gauge invariant term of the form: 
$F_{\mu\nu}dx^{\mu}\wedge dx^{\nu}\wedge*(dx^{\mu}\wedge dx^{\nu})$
as part of the action. However, such a term is zero due to permutation symmetry since 
$W^{\nu\mu}=W^{\mu\nu}
=dx^{\mu}\wedge dx^{\nu}\wedge*(dx^{\mu}\wedge dx^{\nu})
\propto\eta^{\mu\mu}\eta^{\nu\nu}dx^{0}\wedge dx^{1}\wedge dx^{2}\wedge dx^{3}$;
thus, the anti-symmetric $F$ and the symmetric $W$ contract to zero
($F_{\mu\nu}W^{\mu\nu}=0$).

\subsection{Justifying the Einstein--Hilbert--Cartan Action \label{Einstein-Hilbert-Cartan Action}}

For our next example, let us look at the terms related to the matter
equations that involve gravity. There are two possible choices of
matter equation. The first one is the geodesic equation $d\vec{v}/d\tau=\vec{v}\cdot\Gamma\cdot\vec{v}$
where $\Gamma$ is considered as a connection 1-form that transforms
in the usual way $\Gamma\rightarrow\Gamma+\partial g$ under coordinate
transformations by the group element $g$. This type of transformation,
however, is not a ``good'' symmetry since restricting the gauge
transformation $\Gamma\rightarrow\Gamma+\Sigma$ to transformations 
$\Sigma=\partial g$ such that $\vec{v}\cdot\Sigma\cdot\vec{v}=0$,
would mean to select a subset of coordinate systems, inertial systems,
for which the action $S$ is well defined and satisfies $S\left[\Gamma\right]=$
$S\left[\Gamma+\Sigma\right]$. Selecting a specific class of coordinate
systems for the description of a physical phenomenon is not desirable,
so this option shall not be explored any further.

In general, the Euler--Lagrange equations assume a background observer
who defines the coordinate system. For electromagnetism, this is acceptable
since neutral particles are such privileged observers. In gravity,
however, there is no such observer, and the equation for matter should
be relational. Such an equation then is the equation of the geodesic
deviation: $d^{2}\vec{\xi}/d\tau^{2}=\boldsymbol R(v,v)\cdot\vec{\xi}$, where $\boldsymbol R$
is a Lie algebra $(TM)$ valued curvature 2-form $\boldsymbol R=d\Gamma+[\Gamma,\Gamma]$.
A general curvature 2-form is denoted by $F\rightarrow$ $\left(F_{\alpha\beta}\right)_{j}^{i}$.
Here, $\alpha$ and $\beta$ are related to the tangential space $(TM)$
of the base manifold $M$. The $i$ and $j$ are related to the fiber
structure of the bundle over $M$ where the connection $\left(\Gamma_{\alpha}\right)_{j}^{i}$ that defines
$\left(F_{\alpha\beta}\right)_{j}^{i}$ is given. Clearly, the Riemann curvature
tensor $\boldsymbol R$ is a very special curvature because all of its indices
are of the $TM$ type. For that reason, it is possible to contract the
fiber degree of freedom with the base manifold degree of freedom (indices).
Thus, an action linear in $\boldsymbol R$ is possible. In general, one needs
to consider a quadratic action ($F_{\alpha\beta j}^{i}\wedge*F_{\alpha\beta i}^{j}$), i.e. trace of $F\wedge*F$.

Using the symmetries of the Riemann  curvature tensor $\boldsymbol R$ 
($R_{\alpha\beta,\gamma\rho}=-R_{\beta\alpha,\gamma\rho}=-R_{\alpha\beta,\rho\gamma}=R_{\gamma\rho,\alpha\beta}$),
one has two possible expressions that can be proportional to the volume
form $\Omega$. The first expression is possible in all dimensions
and can be denoted by $\boldsymbol R^{*}$, which means that a Hodge dual operation
has been applied to the second pair of indices ($R_{\alpha\beta,*(\gamma\rho)}$).
The $\boldsymbol R^{*}$ action seems to be related to the Cartan--Einstein action
for gravitation $S\left[R\right]=\int R_{\alpha\beta}\wedge*(dx^{\alpha}\wedge dx^{\beta})$
\cite{Adak_et_al_2001}. The other expression is only possible in
a four-dimensional space-time and involves full anti-symmetrization
of $R$ ($R_{\alpha[\beta,\gamma]\rho)}$) denoted by $R^{\wedge}$.
However, the fully anti-symmetric tensor $\boldsymbol R^{\wedge}$ is identically zero
due to symmetry considerations related to the permutation group \cite{Bekaert_and_Boulanger2006}. 
Since the symmetries of the  equation of the geodesic deviation are encoded 
in the Riemann curvature tensor $\boldsymbol R$, then once again one arrives 
at the unique Einstein--Hilbert--Cartan action for gravity based on $\boldsymbol R$. 

\section{Conclusions and Discussion   \label{Conclusions}}

In conclusion, the discussion in this paper showed the potency
of the {principle of reparametrization invariance}
when realized via the {canonical-form of the first-order homogeneous
Lagrangians} in the {velocity} or  {generalized velocity}
by using the {principle of one-to-one correspondence between
an interaction field and its source} to justify the fundamental interaction
fields for the classical long-range forces via the geometrical concepts
of embedding of manifolds as well as the natural differential structures
over manifolds. In summary, the structure of the matter Lagrangian
($L$) for extended objects, and in particular the point particle,
have been discussed. Imposing reparametrization invariance of the
Lagrangian based action $S=\int_{E}~L(x,\boldsymbol \omega)$ 
naturally leads to a first-order homogeneous Lagrangian. 
{In its canonical form, the Lagrangian $L$ contains
electromagnetic and gravitational interactions, 
as well as interactions that are not yet experimentally discovered 
but may be detected as a result of various efforts to address 
current discordances present between the different cosmological \mbox{probes
\cite{Di Valentino et al.(2020)}}.}

The fields $A_\mu(x)$ and $g_{\mu\nu}(x)$ associated with 
$n=1$ and $n=2$ homogeneous Lagrangians built from 
monomials in the velocities $S_n(v,\dots ,v)$ are clearly related to electromagnetic 
and gravitational interactions. Especially, if one recognizes that 
the gauge symmetry of these interaction fields are encoded in the 
2-forms $F$ and $\boldsymbol R$ that naturally appear in
the corresponding equations of motion -- 
the Euler--Lagrange equation that corresponds to the Lorentz force 
$d\vec{v}/d\tau=q F\vec{v}$ for charged particles and 
the equation of the geodesic deviations for massive particles  
$d^{2}\vec{\xi}/d\tau^{2}=\boldsymbol R(v,v)\vec{\xi}$.

If one extrapolates from the strengths of the two known classical
long-range interactions, then it is natural to expect that the new
terms in $L$ should be important, if present in nature at all, at
big cosmological scales, such as those relevant to the dynamics of galactic and galactic clusters. 
Thus, perhaps relevant to the dark matter and dark energy phenomena \cite{Di Valentino et al.(2020)}.
Furthermore, the pathological behavior (\ref{unusual behavior})
discussed for the simplest model of $S_{n}(v,\ldots v)$ fields when
$n>2$, {may be relevant to the inflation processes in
the early universe \cite{Di Valentino et al.(2020)}}. 
At microscopic scales such $n>2$,  fields may be useful in 
justifying the interactions in the standard model of elementary particles 
upon suitable quantization that recovers the Dirac equation but with 
additional interactions beyond electromagnetism and gravitation. 

If one is going to study the new interaction fields $S_{n}(v,\ldots v),n>2$,
then the guiding principles for writing field Lagrangians, as discussed
in the examples of electromagnetism and gravity (Section \ref{Field Lagrangians}),
may be a useful starting point. Furthermore, it may be useful
to apply the outlined constructions to gravity by considering
it as a $3$-brane in a $10$-dimensional target space ($g_{\alpha\beta}:M\rightarrow S_{2}M$)
and to compare it with the $10D$ supergravity.

{If such $S_{n}(v,\ldots v)$ related forces are not present
in nature then one needs to understand why nature is not taking advantage
of such possibilities.} The choice of the canonical Lagrangian is
based on the assumption of one-to-one correspondence between interaction
fields and the type of  sources. {If one can show that any
first-order homogeneous function can be written in the canonical form
proposed, then this would be a significant step towards our understanding
of the fundamental interactions in nature}, especially if one can
show that only $n=1$ and $n=2$ effective terms are needed. Note that an equivalent
expression can be considered as well: $L=A_{\alpha}(\vec{x},\vec{v})v^{\alpha}$.
This expression is simpler and is concerned with the structure of
the homogeneous functions of order zero $A_{\alpha}(\vec{x},\vec{v})$.
In any case, understanding the structure of the homogeneous functions
of any order seems to be an important mathematical problem with significant
implications for physics.

\section{Examples and Exercises \label{Exercises}}
\begin{enumerate}
\item Show that $g_{\mu\nu}v^{\mu}v^{\nu}=constant$ along the trajectory
of a particle is a necessary and sufficient condition for Euler--Lagrange
equations corresponding to $S_{1}$ (\ref{S1}) and $S_{2}$ \mbox{(\ref{S2})}
to be equivalent to each other and to the geodesic equation (\ref{geodesic equations}). 
In the traditional case of velocity independent metric see \cite{Randers41}; 
\item Show that for any Lagrangian $L(x,v)$ that is a homogeneous function in
the velocity $\vec{v}$ of order $n\neq1$ the corresponding Hamiltonian
function $h=v^{\alpha}\left(\frac{\partial L}{\partial v^{\alpha}}\right)-L$
is proportional to the Lagrangian, that is, $h=(n-1)L$;
\item Show that any time independent Lagrangian $L(\vec{x},\vec{v})$,
which is a homogeneous function in velocity $\vec{v}$ of order $n\neq1$,
is an integral of the motion with respect to the corresponding Euler--Lagrange
equations for $L$;
\item Consider a Lagrangian that is a constant of the motion; that is, $dL/d\tau=0$.
Show that any solution of the Euler--Lagrange equations for $L$ is
also a solution for $\tilde{L}=f\left(L\right)$ under certain minor
and reasonable requirements on $f$, such as $\tilde{L}=f\left(L\right)\neq0$
and $\tilde{L}'=f'\neq0$;
\item Show that if $v^{0}=dt/d\tau$ is well behaved ($v^{0}\neq0$ over
the duration of the process studied) then the Euler--Lagrange equations
for the reparametrization-invariant Lagrangian $L(x^{\mu},v^{\mu})=L(x^{\mu},v^{i}/v^{0})v^{0}$,
where $i=1,\ldots,n$, $\mu=0,1,\ldots,n$ and $x^{0}=t,v^{i}=dx^{i}/d\tau,v^{0}=dt/d\tau$,
are equivalent to the Euler--Lagrange equations for coordinate-time
parametrization ($\tau=t$) choice for $L(t,x^{i},dx^{i}/dt)$. Hint:
Use that $L(x^{\mu},v^{i}/v^{0})$ is a zero-order homogeneous function
with respect to $v^{\mu}$ and notice the relationship between the
Hamiltonian function $h$ for the initial Lagrangian $L(t,x^{i},dx^{i}/dt)$
and the generalized momentum $p_{0}=\partial L/\partial v^{0}$ for
the reparametrization-invariant Lagrangian $L(x^{\mu},v^{\mu})=L(x^{\mu},v^{i}/v^{0})v^{0}$;
\item Show that $\sum_{\beta}v^{\beta}\frac{\partial^{2}L}{\partial v^{\alpha}\partial v^{\beta}}=0$
if $L$ is first-order homogeneous Lagrangian. Thus, $\det\left(\frac{\partial^{2}L}{\partial v^{\alpha}\partial v^{\beta}}\right)=0$,
since in an extended space-time one usually expects $v^{0}\neq0$'
\item Consider the constraint $\sqrt{g_{\alpha\beta}v^{\alpha}v^{\beta}}=1$
implemented via a Lagrangian multiplier $\chi$ in the Lagrangian
$L=qA_{\alpha}v^{\alpha}+(m+\chi)g_{\alpha\beta}v^{\alpha}v^{\beta}-\chi$.
Show that the value of $\chi$ is required to be $\chi=-m/2$ if $L=qA_{\alpha}v^{\alpha}+m\sqrt{g_{\alpha\beta}(x)v^{\alpha}v^{\beta}}$
and $L=qA_{\alpha}v^{\alpha}+(m+\chi)g_{\alpha\beta}v^{\alpha}v^{\beta}-\chi$
are to result in the same Euler--Lagrange equations;
\item Show that the function $S_{n}(r,w,u)$ defined in Equation (\ref{S_=00007Bn=00007D(r,w,u)})
is an integral of motion for the equations given by (\ref{u-equation})
and (\ref{w-equation});

\item Consider the Lagrangian 
$L=m\sqrt{\eta_{\alpha\beta}v^{\alpha}v^{\beta}}+\kappa\sqrt[n]{S_{n}\left(\vec{v},\dots ,\vec{v}\right)}$,
where $\eta_{\alpha\beta}=(1,-1,\ldots,-1)$ is the Lorentz invariant
metric tensor. Show that in the non-relativistic limit ($v\rightarrow0$),
the Euler--Lagrange equations for the acceleration $\frac{dv}{d\tau}$
are the same up to $O(v^{2})$ terms whether $L=const$ or $\sqrt{\eta_{\alpha\beta}v^{\alpha}v^{\beta}}=const$
parametrization is imposed. Thus the non-relativistic limit cannot
distinguish these two choices of trajectory parametrization;

\item Show that solutions of the Euler--Lagrange equations for $L=v^{\mu}A_{\mu}(x,v),$
where $x$ is a space-time coordinate and $v^{\mu}$ is a world-line
velocity vector (4-vector for 3+1 space-time), satisfy 
$\frac{d}{d\lambda}\left(v^{\alpha}g_{\alpha\beta}\left(x,v\right)v^{\beta}\right)=0$
for the velocity dependent metric 
$g_{\alpha\beta}\left(x,v\right)=\frac{1}{2}(A_{\alpha/\beta}\left(x,v\right)+A_{\beta/\alpha}\left(x,v\right))$
with $A_{\beta/\alpha}\left(x,v\right)$ being a partial derivative
with respect to $v^{\alpha}$ of $A_{\beta}\left(x,v\right)$;

\item \label{EMandGravityOnly} Choose a specific Lagrangian $\tilde{L}(x,v)$ that is a homogeneous function of first order in $v$,
then consider the Lagrangian $L=v^{\mu}A_{\mu}(x)+\sqrt{g_{\alpha\beta}v^{\alpha}v^{\beta}}$ where
the fields $g_{\alpha\beta}(x)$ and $A_{\mu}(x)$ are defined via the following expressions:
$A_{\mu}(x)=\frac{1}{2}(\tilde{L}(x,v)-\tilde{L}(x,-v))_{/\mu} \left|_{v=(1,\overrightarrow{0})}\right. $ and
$g_{\alpha\beta}(x)=\frac{1}{4}(\tilde{L}(x,v)+\tilde{L}(x,-v))^2_{/\alpha /\beta} \left|_{v=(1,\overrightarrow{0})}\right. $.
Compare the corresponding Euler--Lagrange equations of motions for $\tilde{L}$ and $L$. 
At what order $k$ of $O(v^{k})$ are the differences?

\end{enumerate}

\vspace{6pt} 




\authorcontributions{{For research articles with several authors, a short paragraph specifying their individual contributions must} be provided. The following statements should be used ``Conceptualization, X.X. and Y.Y.; methodology, X.X.; software, X.X.; validation, X.X., Y.Y. and Z.Z.; formal analysis, X.X.; investigation, X.X.; resources, X.X.; data curation, X.X.; writing---original draft preparation, X.X.; writing---review and editing, X.X.; visualization, X.X.; supervision, X.X.; project administration, X.X.; funding acquisition, Y.Y. All authors have read and agreed to the published version of the manuscript.'', please turn to the  \href{http://img.mdpi.org/data/contributor-role-instruction.pdf}{CRediT taxonomy} for the term explanation. Authorship must be limited to those who have contributed substantially to the work~reported.}
\funding{“This research received no external funding” or “This research was funded by NAME OF FUNDER, grant number XXX” and “The APC was funded by XXX”. Check carefully that the details given are accurate and use the standard spelling of funding agency names at https://search.crossref.org/funding, any errors may affect your future funding.}

\acknowledgments{V. G. is extremely grateful to his wife and daughters for their understanding and family support. Past preliminary research was performed at Louisiana State University and Lawrence Livermore National Laboratory. V.G. acknowledges past discussions on the topic with Dr Carlos Castro Perelman and Professors R. Haymaker, A. R. P. Rau, P. Kirk, J. Pullin, R. O'Connell, C. Torre, J. Baez, P. Al. Nikolov, E. M. Prodanov, G. Dunne, L. I. Gould, and D. Singleton.}

\conflictsofinterest{``The authors declare no conflict of interest.'' Authors must identify and declare any personal circumstances or interest that may be perceived as inappropriately influencing the representation or interpretation of reported research results. Any role of the funders in the design of the study; in the collection, analyses or interpretation of data; in the writing of the manuscript, or in the decision to publish the results must be declared in this section. If there is no role, please state ``The funders had no role in the design of the study; in the collection, analyses, or interpretation of data; in the writing of the manuscript, or in the decision to publish the~results''.} 

\end{paracol}
\reftitle{References} 



\end{document}